\documentclass[12pt]{iopart}
\usepackage{graphicx} 

\def\prl{{\em Phys. Rev. Lett. }}
\def\prc{{\em Phys. Rev. {\bf C} }}
\def\prd{{\em Phys. Rev. {\bf D} }}

\def\jpg{{\em J. Phys. {\bf G} }}

\def\npa{{\em Nucl. Phys. {\bf A} }}
\def\npb{{\em Nucl. Phys. {\bf B} }}
\def\epjc{{\em Eur. Phys. J. {\bf C} }}
\def\epja{{\em Eur. Phys. J. {\bf A} }}
\def\plb{{\em Phys. Lett. {\bf B} }}

\def\zpc{{\em Z. Phys. {\bf C} }}

\def\app{{\em Acta Physica Polonica {\bf B} }}
\def\rmp{{\em Rev. of Mod. Phys. }}
\def\ijmpa{{\em Int. J. Mod. Phys. {\bf A} }}
\def\ijmpe{{\em Int. J. Mod. Phys. {\bf E} }}
\begin{document}
\title{Nuclear suppression of heavy quark production at \
 forward rapidities in relativistic heavy ion collisions}
\title[Nuclear suppression of heavy quark production..]{}

\author{Umme Jamil and Dinesh K. Srivastava}
\address{Variable Energy Cyclotron Centre, 1/AF Bidhan Nagar, Kolkata 700064, 
India}

\begin{abstract}
We calculate nuclear suppression $R_{AA}$ of heavy 
quarks produced from the initial fusion of partons 
in  nucleus-nucleus collisions at RHIC and LHC energies. 
We take the shadowing  as well 
as the energy loss suffered by them while passing through 
Quark Gluon Plasma into account. We obtain results for charm and bottom
quarks at several rapidities using different mechanisms for
energy loss, to see if we can distinguish between them.

\end{abstract}
\section{Introduction}

\ \ \ \ \ \ 
The heavy ion collision experiments at RHIC and LHC are designed with 
a hope to explore the existence of a new form of matter known as 
Quark Gluon Plasma (QGP) and to explore its properties. 
The estimation for the energy density~\cite{br1, st1, ph1, phe1}
 attained in these 
collisions using the Bjorken formula~\cite{info1} is well beyond the
energy densities where QGP is expected to be formed. 
The temperatures reached at RHIC, as revealed from several studies (see e.g.,
~\cite{de,phenix_gam,fms} for a compilation)
are also much larger than the values provided by Lattice 
QCD calculations 
for the critical temperature for a transition to QGP~\cite{info2}.
Strong confirmation of the formation of the QGP is given by
observation of a large elliptic flow~\cite{elliptic}, jet-quenching~\cite{jet},
and the recombination of partons as the mechanism of production of
hadrons at intermediate transverse momenta~\cite{recomb}.  
 Still higher temperatures are likely to be reached at
LHC.

Heavy quarks are produced from the initial fusion
of gluons ($gg\rightarrow Q\overline{Q}$) or light quarks 
($q\overline{q}\rightarrow Q\overline{Q}$). This pair would be produced at
$\tau \approx 1/2M_Q \ll $ 0.1 fm/c.
Their large mass ensures that their production can be treated using pQCD 
and that it is nearly negligible at later times. 
These will traverse the QGP, colliding with quarks and 
gluons and radiating gluons before appearing as charm or bottom
mesons or baryons. Thus the final spectra for these hadrons would contain
information about the energy loss suffered by the heavy quarks.
The unique importance of heavy quarks as probes of QGP lies in their
large mass. This leads to a considerably reduced production of heavy quarks
in comparison to light quarks and gluons which are produced copiously. The
strong interaction during the collision conserves flavour.  Even at the LHC, 
the reverse process of annihilation of heavy quarks 
($Q\overline{Q}\rightarrow gg$, etc.) can be safely ignored. Thus heavy quarks
and in turn charm and bottom hadrons will stand out in the back-ground of a
multitude of light hadrons, and one can in principle track them.

Along with other reasons, an 
interest in the study of energy loss of heavy quarks was triggered by the
large back-ground that correlated charm or bottom decay 
provides~\cite {ramona,sean} to the thermal
dileptons which have been considered a signature of the formation of QGP
for a long time~\cite {shuryak1, hwa, kkmr, kkmm, ssc}. It was pointed out by 
Shuryak~\cite{shuryak}, 
Lin {\it et al.}~\cite{lin, lin1}, Kampfer {\it et al.}~\cite{kampfer}, and 
Mustafa {\it et al.}~\cite{mus1, mus2}
 that the correlated charm and bottom decay 
could be suppressed if the energy loss suffered by heavy quarks before
they form D or B mesons was accounted for. Since then several attempts have 
been made to estimate the energy loss of heavy quarks as they proceed through
QGP. Of course the possibility to identify the vertex of D or B meson decay 
will further enrich this study.

These results have been put to a rigorous
 test by the measurement of single electrons
from heavy ion collisions at RHIC which show a clear evidence for 
the loss of energy by the heavy quarks~\cite{info3}. Their possible 
thermalization is also indicated
 by the elliptic flow that they show~\cite{ellip}.

As indicated earlier, the temperature likely to be reached at LHC in
collision of heavy nuclei could be even larger and thus this energy loss
will play a more significant role. The opening of a much
 wider window in rapidity at LHC is also likely to provide widely 
differing media at different rapidities through which the heavy quarks
would propagate.

Thus a  valuable test of various theories for energy loss
 suffered by heavy quarks can be performed by studying it at RHIC and LHC and
at different rapidities.

We study these effects in terms of nuclear modification factor $R_{AA}$ for
heavy quarks. In these initial studies we calculate the average energy
loss suffered by them as they pass through 
the QGP and compare the resulting $p_T$ 
distribution with the same for proton-proton (pp)
 collisions to get $R_{AA}$. Since the mass 
of charm or bottom quarks is quite large, the $p_T$ distribution of these 
quarks will closely reflect the $p_T$ distribution of D or B mesons.

We employ a local fluid approximation~\cite{ramona,sean,sourav} in
order to picture the medium at larger rapidities. We shall come back to this
later.

The paper is organized as follows. As we need to compare the spectra of the
heavy quarks from relativistic heavy ion collisions with those for pp
 collisions, as a first step we study the heavy quark production in LO pQCD
and compare our results with a NLO pQCD calculation. We find that single
quark distribution calculated using LO pQCD supplemented with a K-factor
adequately reproduces the NLO results as well as the available experimental
data. Next we estimate the average energy loss suffered by heavy quarks of a 
given energy using various mechanisms discussed in the literature.

Finally we perform a Monte Carlo calculation to obtain the average 
change in the
transverse momentum spectra of heavy quarks for nucleus-nucleus collisions
and get $R_{AA}$ as a function of $p_T$ for different 
rapidities. We add that this work is not intended as a complete review and the 
readers may see Ref.~\cite{mus, guy}, for other treatments.

More detailed calculations where the consequences of energy loss of heavy quarks
on the correlated charm or bottom decay and modification of the back-to-back
correlation of heavy quarks are discussed, will be published shortly.

\section{Heavy quark production in pp collisions}

\ \ \ \ \ \
At lowest order in pQCD, heavy quarks in pp collisions are produced by
fusion of gluons ($gg \rightarrow Q\overline{Q}$) or light quarks 
($q \overline{q} \rightarrow Q\overline{Q}$)~\cite{comb}. 
The so-called flavour excitation process ($qQ \rightarrow qQ$ and 
$gQ \rightarrow gQ$) is now known to be suppressed when the NLO processes
are taken into account~\cite{lin-gussy, collins, ellis}. In addition, Brodsky 
{\it et al.}~\cite{bro1, bro2} have shown that the total contribution 
of intrinsic charm in the midrapidity region is small even though most of the 
heavy quarks are produced in this region.

The cross-section  for the production of heavy quarks from pp collisions at 
lowest order is
given by~\cite{comb, cs1}:
 
\begin{eqnarray}
\frac{d \sigma}{dy_1\,dy_2\,dp_T}=
2 x_1x_2 p_T\sum\limits_{ij}& &[f_i^{(1)}(x_1, Q^2)f_j^{(2)}
(x_2, Q^2)
\hat\sigma_{ij}(\hat{s}, \hat{t}, \hat{u})+\nonumber\\
& &f_j^{(1)}(x_1, Q^2)f_i^{(2)}(x_2, Q^2)
\hat\sigma_{ij}(\hat{s}, \hat{t}, \hat{u})]/(1+\delta_{ij}).
\label{ds}
\end{eqnarray}
In the above equation, i and j are the interacting partons, 
$f_i^{(1)}$ and $f_j^{(2)}$ are the partonic structure functions and 
$x_1$ and $x_2$ are the fractional momenta of the interacting hadrons
 carried by the partons i and j. The relation between $p_T$ and fractional 
momentum $x_1$ or $x_2$ through their respective rapidities can be written as
\begin{eqnarray}
x_{1}=\frac{m_T}{\sqrt s}\ (e^{y_1} \ + \ e^{y_2} ), \ 
x_{2}=\frac{m_T}{\sqrt s}\ (e^{- y_1} \ + \ e^{-y_2} ),
\label{x1}
\end{eqnarray}
where $m_T$ is the transverse mass, $\sqrt{M^2\,+\,p_T^2}$, 
of the produced heavy quark.
The function $\hat\sigma\,=\,d\sigma/dt$, the short range 
subprocess for the heavy quark production is defined as: 
\begin{eqnarray}
\frac{d\sigma}{dt}=\frac{1}{16\pi\hat{s}^2}\,|\mathcal{M}|^2.
\end{eqnarray}
 $|\mathcal{M}|^2$ for the 
heavy quark production processes $gg$$\rightarrow$$Q\bar{Q}$ and
 $q\bar{q}$$\rightarrow$$Q\bar{Q}$
 are expressed through the mass of the heavy 
quark and Mandelstam 
variables $\hat{s}$,  $\hat{t}$, and $\hat{u}$ as

\begin{eqnarray}
\bigg|\mathcal{M}\bigg|^2_{(gg\rightarrow Q\bar{Q})}={\pi^2 \alpha_s^2}
& &\left[\frac{12}{\hat{s}^2}\left(M^2-\hat{t}\right)\left(M^2-\hat{u}\right)
\right.\nonumber\\
& &\,+\,\frac{8}{3}\frac{\left(M^2-\hat{t}\right)\left(M^2-\hat{u}\right)
-2M^2\left(M^2+\hat{t}\right)}
{\left(M^2-\hat{t}\right)^2}\nonumber\\
& &\,+\,\frac{8}{3}\frac{\left(M^2-\hat{t}\right)\left(M^2-\hat{u}\right)
-2M^2\left(M^2+\hat{u}\right)}
{\left(M^2-\hat{u}\right)^2}\nonumber\\
& &\,-\,\frac{2M^2\left(\hat{s}-4M^2\right)}{3
\left(M^2-\hat{t}\right)\left(M^2-\hat{u}\right)}\nonumber\\
& &\,-\,6\frac{\left(M^2-\hat{t}\right)\left(M^2-\hat{u}\right)+M^2\left
(\hat{u}-\hat{t}\right)}{\hat{s}\left(M^2-\hat{t}\right)}\nonumber\\
& &\left.\,-\,6\frac{\left(M^2-\hat{t}\right)\left(M^2-\hat{u}\right)+M^2
\left(\hat{t}-\hat{u}\right)}{\hat{s}\left(M^2-\hat{u}\right)}\right]
\end{eqnarray}
and
\begin{eqnarray}
\bigg|\mathcal{M}\bigg|^2_{(q\bar{q}\rightarrow Q\bar{Q})}\,=\,\frac{64 \pi^2 
\alpha_s^2}{9}
\left[\frac{\left(M^2-\hat{t}\right)^2+\left(M^2-\hat{u}\right)^2+2M^2\hat{s}}
{\hat{s}^2}\right].
\end{eqnarray}

The running coupling constant $\alpha_s$ at lowest order is 
\begin{eqnarray}  
\alpha_s=
\frac{12\pi}{(33-2N_f)\ln \left(Q^2/\Lambda^2\right)}, 
\end{eqnarray}
where $N_f$\,=\,3 is the number of active flavours and 
$\Lambda\,=\,\Lambda_{\rm QCD}$.
 We use the factorization and renormalization scales
 as $Q^2$\,=\,$m_T^2$. We refer 
the readers to Vogt {\it et al.}~\cite{vogt1} for results on variations of these 
scales.
We also carry out the 
calculation of differential cross section for heavy quarks in 
pp collision at NLO in pQCD using
 the treatment developed by Mangano, Nason, and Ridolfi (MNR-NLO)~\cite{mnr}. 
 All the 
calculations are carried out by neglecting the intrinsic transverse momentum
of the partons. 

The effect of nuclear shadowing 
in high energy nucleus-nucleus collisions 
is well known~\cite{shad1, shad2, shad3, shad4}. 
With the increase of the mass number of the 
nucleus and increasing contribution of terms having small x, the effect 
becomes more pronounced. We introduce the 
shadowing effect in our calculations by using EKS 98 
parameterization~\cite{eks} for nucleon structure functions. We take
 CTEQ4M~\cite{cteq} structure function set for nucleons.

We shall see that x dependence of the shadowing function 
introduces interesting structures in the nuclear modification factor as a 
function of $p_T$, y, and the incident energy, because of the large mass 
of the quarks. 

\begin{figure}[ht]
\begin{center}
\includegraphics[width=14 pc,angle=-90]{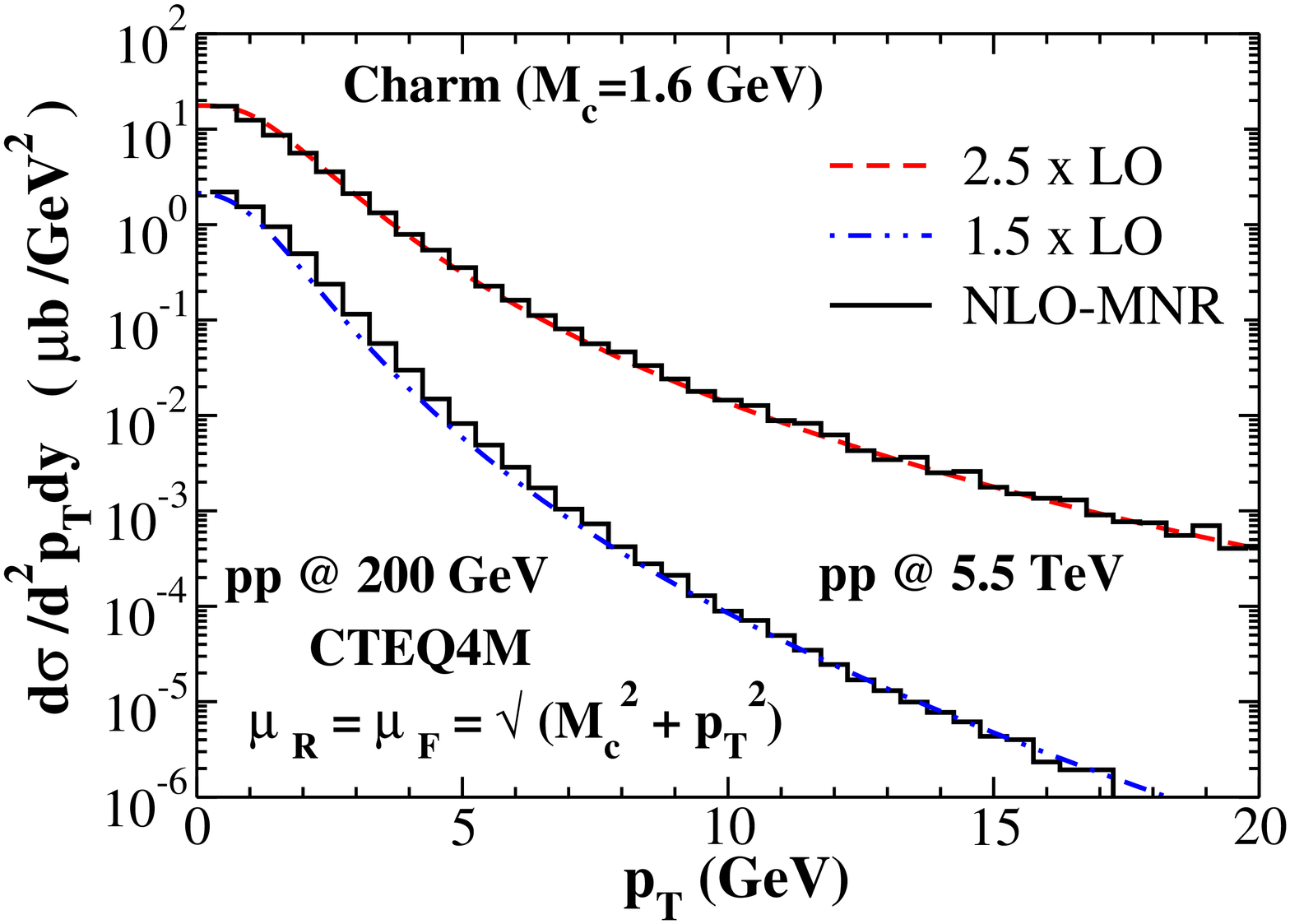}
\includegraphics[width=14 pc,angle=-90]{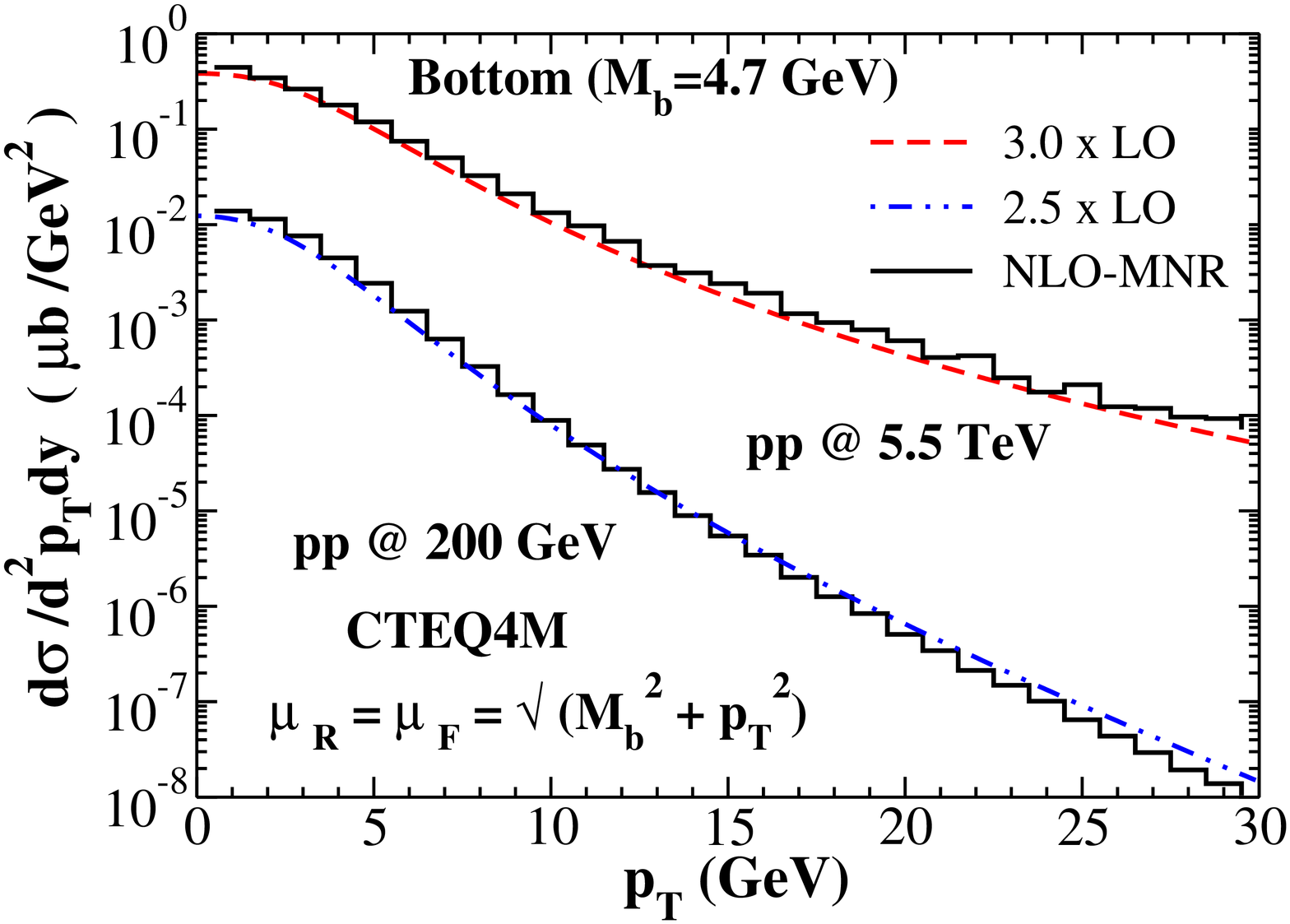}
\caption{\label{2} [Left panel] Comparison of our lowest order 
pQCD results with the 
NLO-MNR calculation
 for charm quark $\left(M_c\,=\,1.6 \,{\rm {GeV}}\right)$ at midrapidity. 
[Right panel]
Same for bottom quark  $\left(M_b\,=\,4.7\, {\rm {GeV}}\right)$ at midrapidity.}
\label{dsdy}
\end{center}
\end{figure}

In Fig.~\ref{dsdy} 
we compare
our results for heavy quark $p_T$ distribution obtained using lowest order 
pQCD for pp collision
 with the results from NLO-MNR calculation at 
midrapidity for charm 
and bottom quarks at RHIC and LHC energies. These comparisons suggest a K 
factor of $\approx$\,1.5\,-\,3 for our lowest 
order calculations for agreement with 
NLO results.
\begin{figure}[ht]
\begin{center}
\includegraphics[width=14 pc,angle=-90]{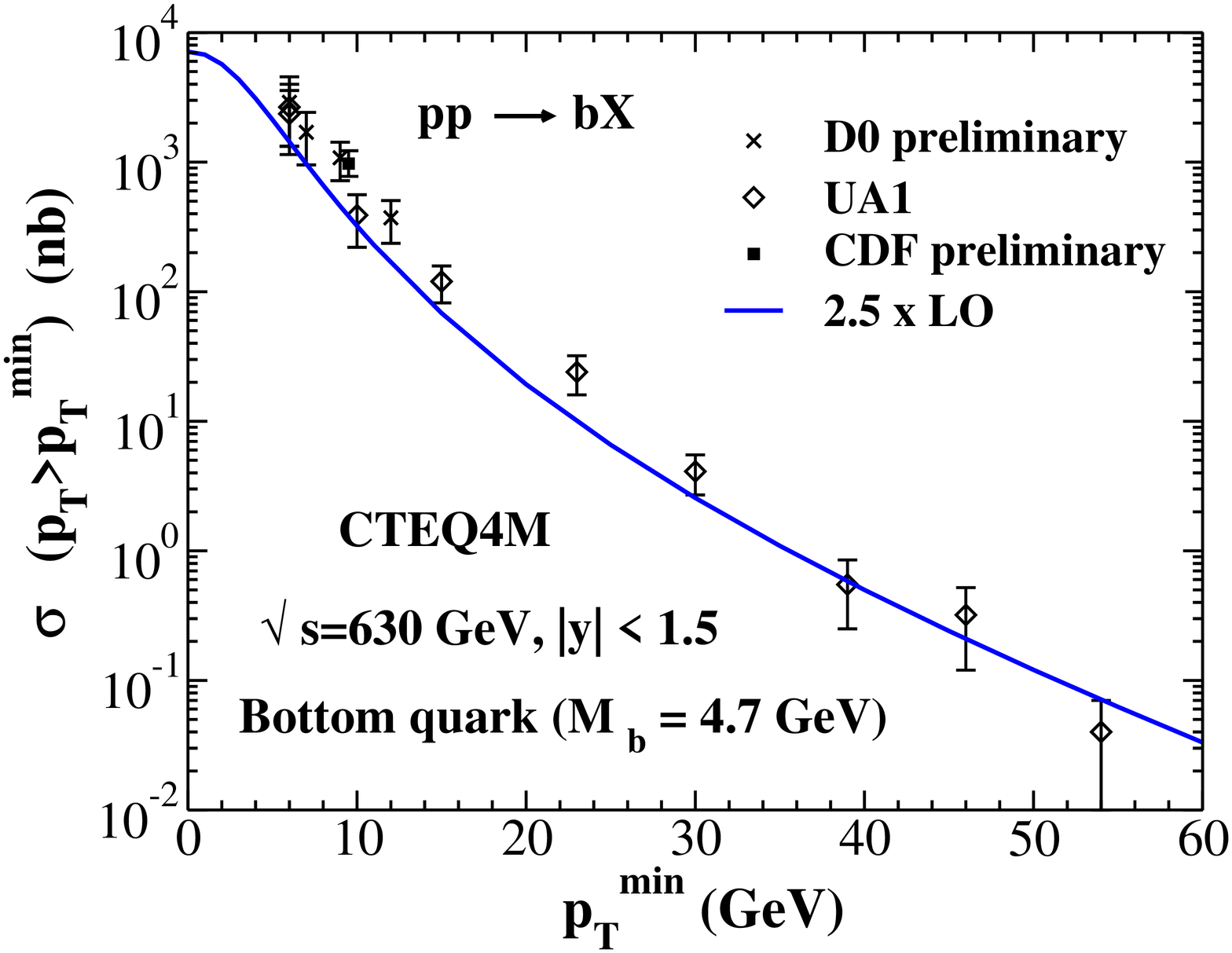}
\includegraphics[width=14 pc,angle=-90]{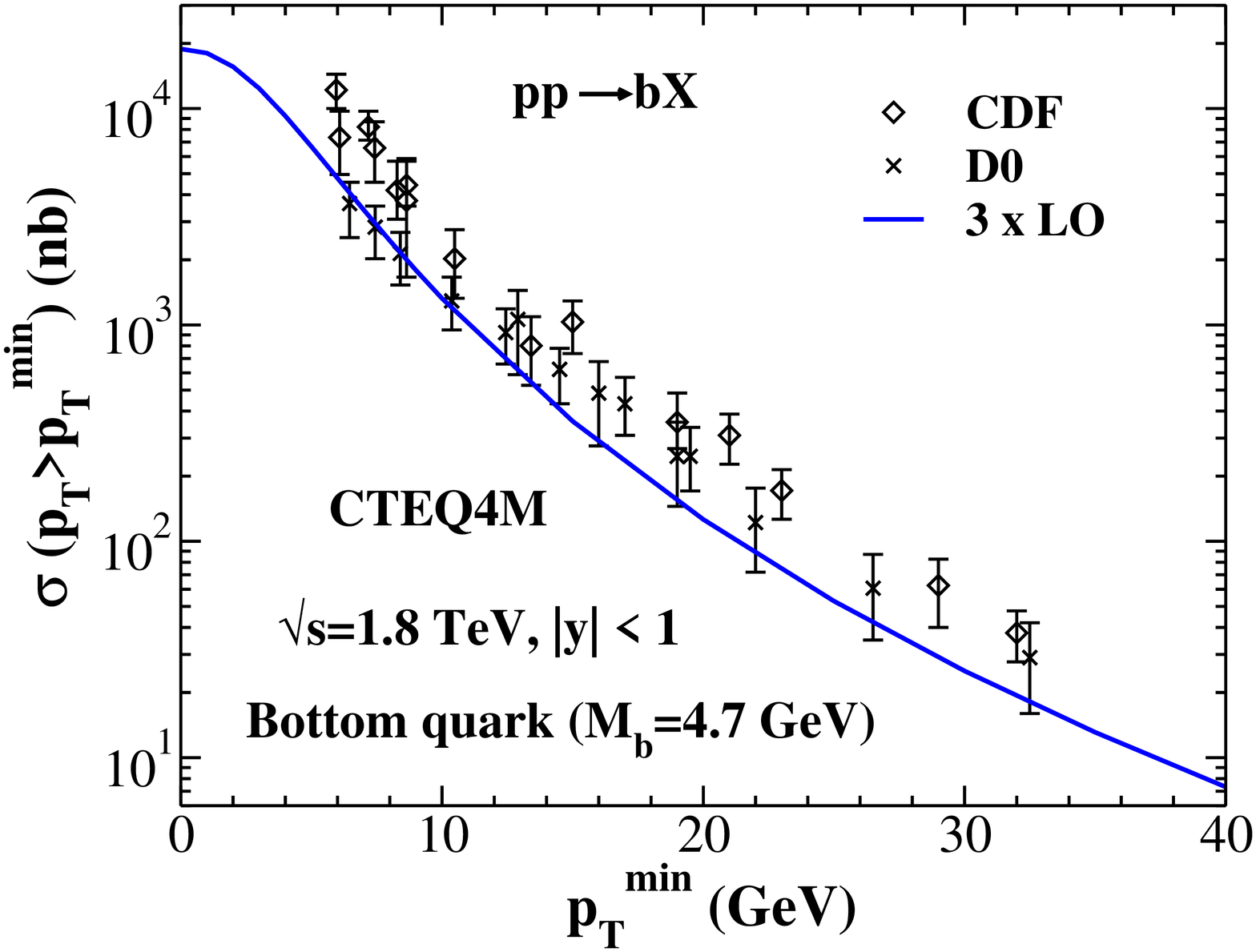}
\caption{\label{2} Comparison of our result for $\sigma\left(p_T\,>\,p_T^{min}
\right)$ for production of bottom quarks with experimental data. }
\label{ptmin}
\end{center}
\end{figure}

In Fig.~\ref{ptmin} we compare the bottom quark production 
cross-section obtained using lowest order pQCD
 in pp collision at 630 and 1800 GeV 
energies with UA1, CDF and $D\O$ data~\cite{edpb1, edpb2}. We find a good 
description of these data using lowest order pQCD with a K-factor.

\begin{figure}[ht]
\begin{center}
\includegraphics[width=14 pc,angle=-90]{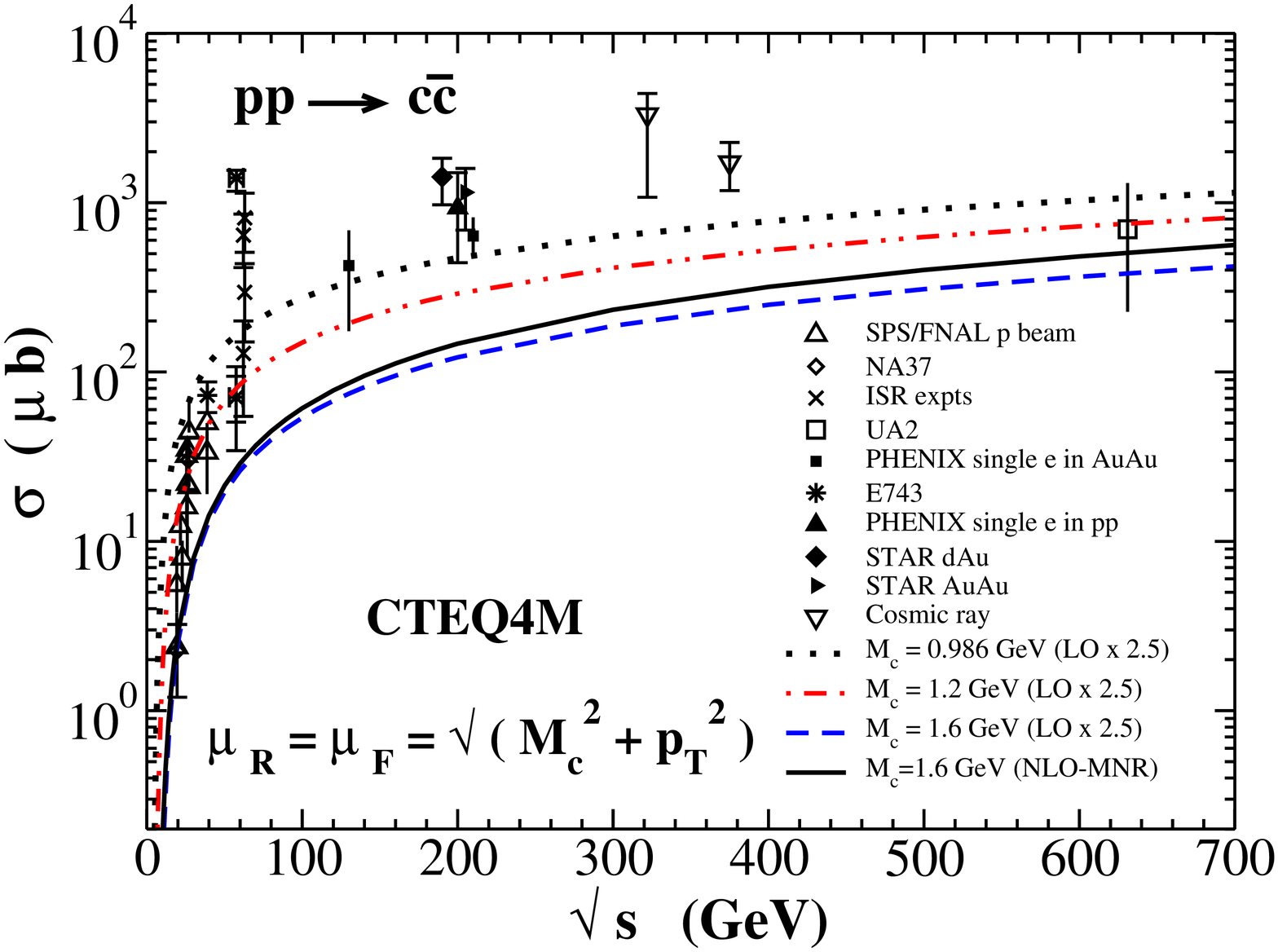}
\includegraphics[width=14 pc,angle=-90]{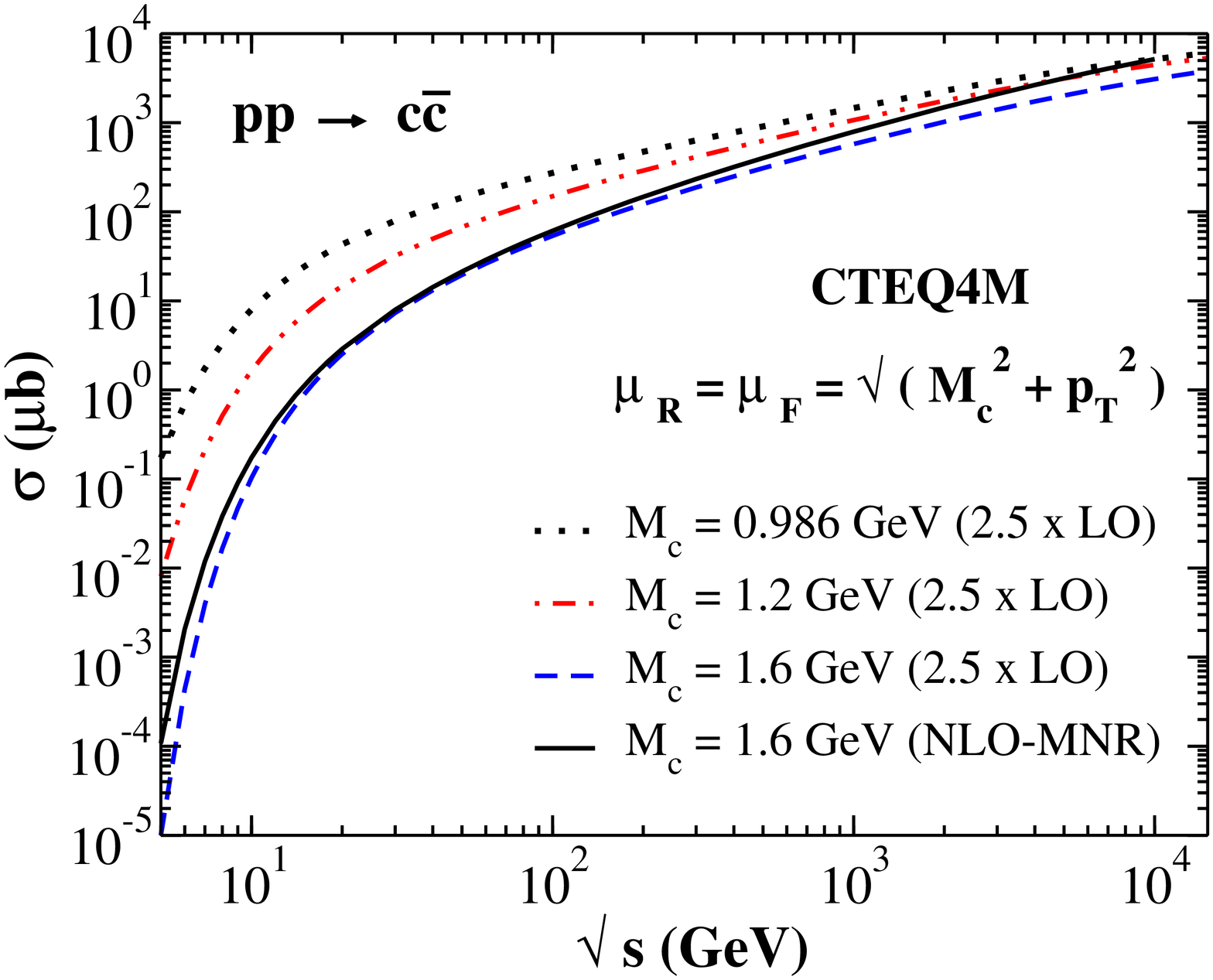}
\caption{\label{2} Total cross-section for pp$\rightarrow$c$\bar c$ 
compared with 
experimental data at varying $\sqrt{s}$.}
\label{csc}
\end{center}
\end{figure}

We calculate the total cross-section for charm quark 
production at lowest order
 for the 
process pp$\rightarrow$$c\bar c$ as a function of $\sqrt{s}$ considering
 the charm quark mass as 
1.2 GeV and 1.6 GeV. We have also included results for $M_c\,(3 \,\rm {GeV})\,
=\,0.986 \,\rm{GeV}$~\cite{mass}, suggested recently. Though the last value 
is not at the pole Q\,=\,$M_c$, 
it may serve as the lower limit to the mass of the 
charm quark. 
In the left panel of Fig.~\ref{csc} we compare these results
 and the results 
from NLO-MNR calculation with the experimental data 
points~\cite{edpb2, edp1, edp2}. Here also 
our lowest order pQCD calculation show a good agreement with experimental 
data points for $M_c$\,=\,1.2 GeV. 
In the right panel of  Fig.~\ref{csc} we compare
  our results with the NLO-MNR calculations up to $\sqrt{s}$\,=\,15000 GeV. 

\begin{figure}[ht]
\begin{center}
\includegraphics[width=16 pc,angle=-90]{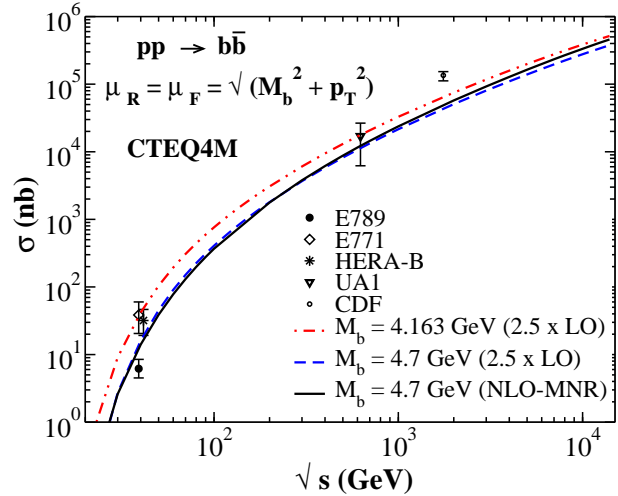}
\caption{\label{2}Total cross-section for pp$\rightarrow b \bar b$ 
compared with 
experimental data varying $\sqrt{s}$.}
\label{csb}
\end{center}
\end{figure}

We also calculate the total cross-section for 
bottom quark production for pp$\rightarrow b \bar b$ as a function 
of $\sqrt{s}$ considering the bottom quark mass as 
4.7 GeV and 4.163 GeV~\cite{mass}. In Fig.~\ref{csb} we compare 
these results and the results obtained from NLO-MNR calculation
 with the experimental data points~\cite{edp3}. 
This comparison is 
quite impressive as at $M_b$\,=\,4.7 GeV our lowest order result 
accurately reproduce the result of NLO-MNR calculation with K\,=\,2.5. 

Thus we see that the $p_T$ distribution and production cross-section for the 
heavy quarks calculated in 
lowest order pQCD and supplemented with a K-factor $\approx$ 2.5 
reproduces the results at NLO for pp collisions. In view of this
 we feel that these distribution 
would be adequate for calculating $R_{AA}$.


\section{The Initial conditions and the Evolution of the Plasma}

The heavy quarks produced at the initial stage pass through the QGP, 
where they loose energy by colliding with quarks and 
gluons and also by radiating gluons. The energy loss will depend upon the 
path-length of the heavy quarks in the plasma, the temperature evolution of the
 plasma, and the energy and mass of the heavy quarks. 

In order to proceed we make 
several simplifying
 assumptions. It is expected that the heavy quarks will loose 
most of their energy when the temperature is still large, i.e. during the 
earliest times after the formation of QGP. During these early times, we can 
neglect the transverse expansion of the plasma. We assume a Gaussian rapidity 
density distribution for the particles produced and further assume that 
the initial rapidity distribution of the quarks and gluons follows this 
distribution closely. This would correspond to an isentropic expansion of the 
plasma at all rapidities, and should be sufficient for our initial studies. 

Thus we assume the rapidity 
distribution of the density of gluons as~\cite{ramona, sean}:
\begin{eqnarray}
 \frac{dN_g}{dy}={\left (\frac{dN_g}{dy}\right )}_0 \ \exp \ \left(-\,y^2/
2\sigma^2\right). 
\end{eqnarray}
 We take ${(\frac{dN_g}{dy})}_0$\,$\approx$\,900 and $\sigma\,=\,3$ 
for Au+Au collisions at RHIC~\cite{mrhic}
 and $\approx$\,3300 and $\sigma\,=\,4$ 
for Pb+Pb collisions at LHC~\cite{mlhc}.

The Bjorken cooling is then assumed to work locally 
at different rapidities, and we 
consider the passage of a heavy quark having rapidity y in a fluid having 
an identical fluid rapidity. 
This approximation, which corresponds to assuming a boost-invariant expansion 
along with a local fluid approximation, has been used earlier in 
literature~\cite{ramona,sean,sourav}. A more complete study would use a 
(3\,+\,1) dimensional hydrodynamics~\cite{hirano, bass}, 
which we plan to use in future publications.

\begin{figure}[ht] 
\begin{center}
\includegraphics[width=16 pc]{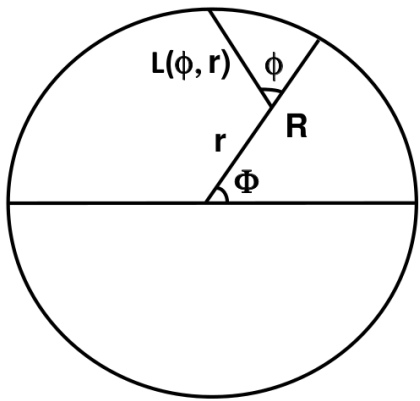}
\caption{\label{2} The distance, L, covered by a heavy quark while passing 
through the QGP. For central collisions the results for $\langle \rm {L}
 \rangle$ 
will not depend on $\Phi$.}
\label{avg}
\end{center}
\end{figure}

We consider a heavy quark produced in a central collision, at the point (r, 
$\Phi$), and moving at an angle $\phi$ with respect to $\hat{\rm {r}}$ in the 
transverse plane. In general the distance covered by the heavy quark before it 
exists the QGP, will vary from 0 to 2R, where R is the radius of the colliding 
nuclei. The distance covered by the heavy quark in the plasma, L, is given by
~\cite{l}:
\begin{eqnarray}
L(\phi,\, {\rm r})=
\sqrt{{\rm R}^2\,-\,{\rm r}^2\sin^2 \phi}\,-\,{\rm r}\cos\phi.
\end{eqnarray}
We can estimate the average distance travelled by the heavy quarks in the 
plasma as: 
\begin{eqnarray}
\langle L \rangle=
\frac{\int\limits_0^{\rm R}{\rm r}\,{\rm dr}
\,\int\limits_0^{2\pi} L(\phi,\, {\rm r}) T_{AA}({\rm r}, b=0)\,{\rm d\,}\phi}
{\int\limits_0^{\rm R}{\rm r}\,{\rm dr}
\,\int\limits_0^{2\pi} T_{AA}({\rm r}, b=0)\,{\rm d\,}\phi}. 
\end{eqnarray}
In the above
 the nuclear overlap function $T_{AA}$(r, b\,=\,0) provides the probability 
of production of heavy quarks in hard binary collisions.
We find that $\langle {\rm {L}} \rangle$ is 5.78 fm 
for Au+Au collisions at RHIC and 6.14 fm for 
Pb+Pb collisions at LHC, and is about 20$\%$ smaller than the radii of the 
colliding nuclei, as the appearance of the nuclear overlap function gives a 
larger weight to the points having smaller r. 

\begin{figure}[ht]
\begin{center}
\includegraphics[width=16 pc,angle=-90]{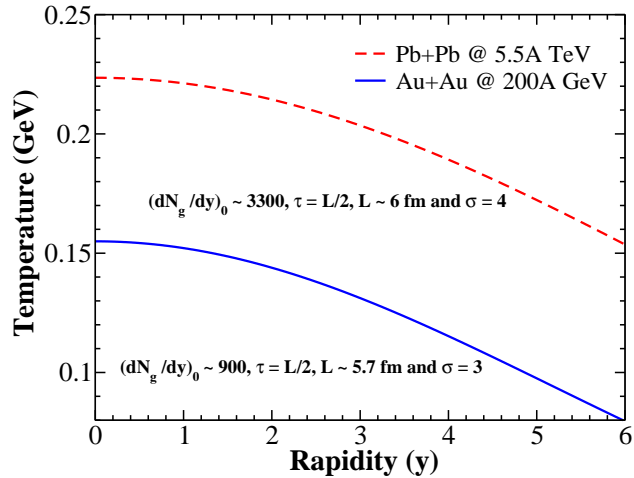}
\caption{\label{2} Variation of temperature of QGP with rapidity 
y at a typical $\tau$.}
\label{tr}
\end{center}
\end{figure}

As the heavy quarks loose most of their energy in interaction with gluons, it
 is enough to consider only the distribution of gluons.
 Their density at the time 
$\tau$ can be 
written as~\cite{dglv2}:

\begin{eqnarray}
\rho\,(\tau) = \frac{1}{\pi\,{\rm R}^2\,\tau}\, \frac{dN_g}{dy}.
\end{eqnarray}
The corresponding temperature~\cite{dglv2}, 
assuming a chemically equilibrated plasma is
\begin{eqnarray}
T\,(\tau) = \left(\frac{\pi^2}{1.202}\, \frac{\rho\,(\tau)}
{(9\,N_f +16)}\right)^{\frac{1}{3}}.
\end{eqnarray}
The rapidity dependence of the temperature of the plasma at a typical 
$\tau\, =\, \langle\,{\rm L}\,\rangle/2$ 
at RHIC and LHC is given in Fig.~\ref{tr}. 

Assuming that the QGP is formed at $\tau_0$\,=\,0.2 fm/c, we estimate the $T_0$ 
at y\,=\,0 for RHIC as 377 MeV and LHC as 555 MeV. More
detailed studies do suggest
a larger formation time of $\approx$ 0.5 fm/c, which will correspondingly
reduce the initial temperatures. This will not affect our results 
as we approximate the expanding and cooling plasma,  with one
at a temperature
determined at $\tau=L_{\rm {eff}}/2$,
 which is much larger (see the following discussion).
Further assuming, Bjorken's 
cooling law, $T^3\, \tau$\,=\,constant, 
this provides that the plasma would cool down 
to the transition or critical temperature $T_c$\,$\approx$\,160 MeV by 
$\tau_c$\,$\approx$\,2.6 fm/c at RHIC and 8 fm/c at LHC. 
One can easily calculate 
the corresponding values at larger rapidities.

Considering the velocity of the quark as $v_T$\,=\,$p_T/m_T$, it would take 
a time $\tau_L$ \,= \,$\langle {\rm L} \rangle/v_T$ to cross the plasma. 
If $\tau_c$ \, $\geq$ \, $\tau_L$ the 
heavy quark would be inside QGP during the entire 
period, $\tau_0$ to $\tau_L$. However,
 if $\tau_c$ \, $\textless$ \, $\tau_L$, only while covering the distance 
$v_T\,\times\, \tau_c$, would the heavy quark be in the QGP phase. 
We further approximate the expanding and cooling plasma with one at a 
temperature of T at $\tau$\, =\, $\langle {\rm L} \rangle_{\rm eff}/2$, where 
$\langle {\rm L} \rangle_{\rm eff}$\,=\,min
$\left[\langle {\rm L} \rangle,\, v_T \,\times \,\tau_c \right]$. 
This procedure 
has been used frequently~\cite{dglv2}.

\section{Mechanisms for Energy Loss}

Next we discuss the energy loss mechanisms that we have included. As mentioned 
above, repeatedly, the heavy quarks loose energy both by collisions as well as 
radiation of gluons. 
A number of formalisms have been proposed for 
the collisional as well as the 
radiative energy loss of heavy quarks in the literature.
 We shall consider the following 
treatments for the collisional energy loss. 

Bjorken~\cite{bjorken} has considered the collisional energy loss of 
light quarks 
as analogous to the energy loss of a charged particle passing through a medium 
and losing energy by ionizing the medium. His expression for massless quarks 
was adapted by Braaten and Thoma to the case of heavy quarks 
[see Eq.~\ref{bjorken}]. We shall continue to label this mechanism as Bjorken 
for clarity. Braaten and Thoma (BT)~\cite{bt1, mus2}
 also modified the expression for the energy 
loss suffered by muons while traversing QED plasma, to obtain the collisional
 energy loss of a heavy quark as it passes through the QGP 
[see Eqs.~\ref{bt1} and ~\ref{bt2} in the Appendix A]. 
These results are valid for 
collisions where the momentum transfer q\,$<<$\,E, where E is the energy of the 
heavy quark. 
Peigne and Peshier (PP)~\cite{pp} have improved this treatment by including the 
u-channel, which becomes important for large energies [see Eq.~\ref{pp} 
in the Appendix A]. 
\begin{figure}[ht]
\begin{center}
\includegraphics[width=30 pc,angle=-90]{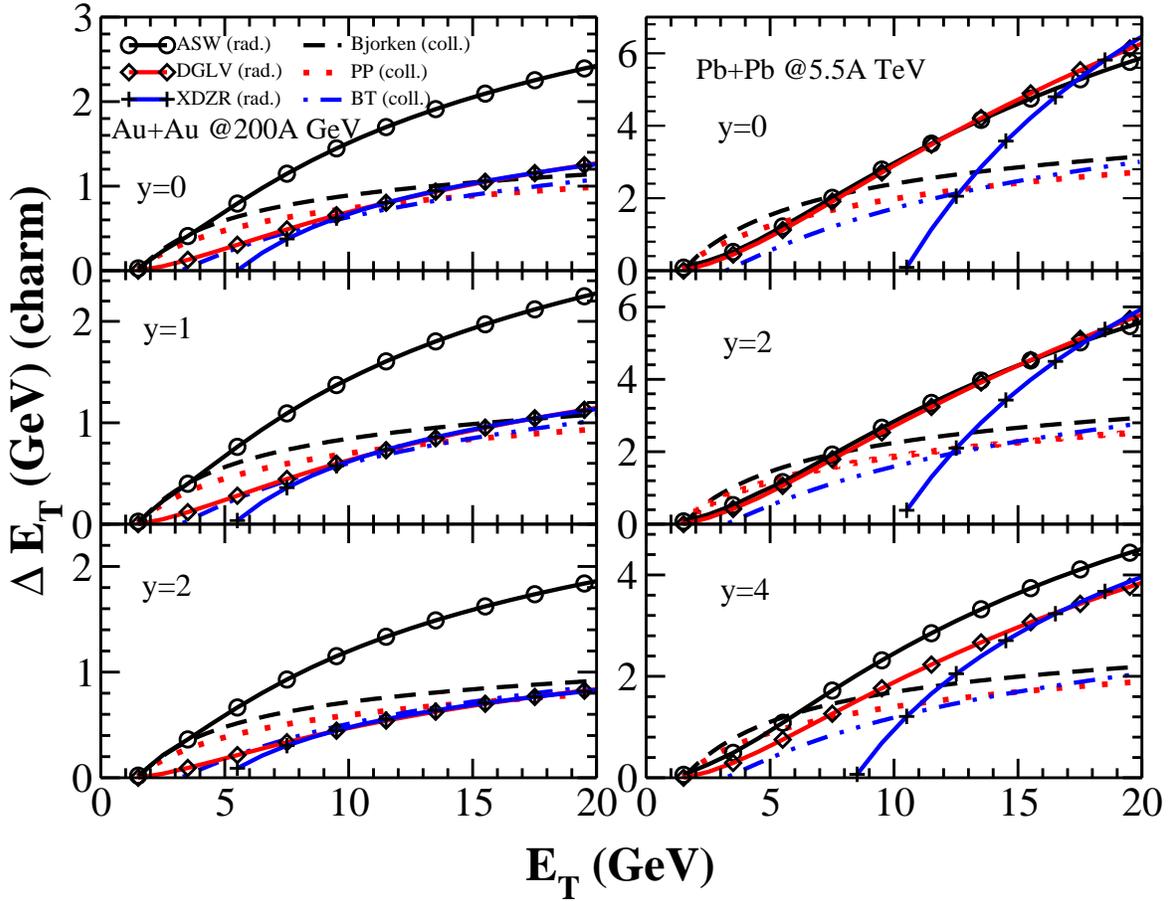}
\caption{\label{2} Collisional (dotted lines) and radiative (solid lines)
energy loss suffered by a charm 
quark while passing through the QGP}
\label{eloss-c}
\end{center}
\end{figure}

\begin{figure}[ht]
\begin{center}
\includegraphics[width=30 pc,angle=-90]{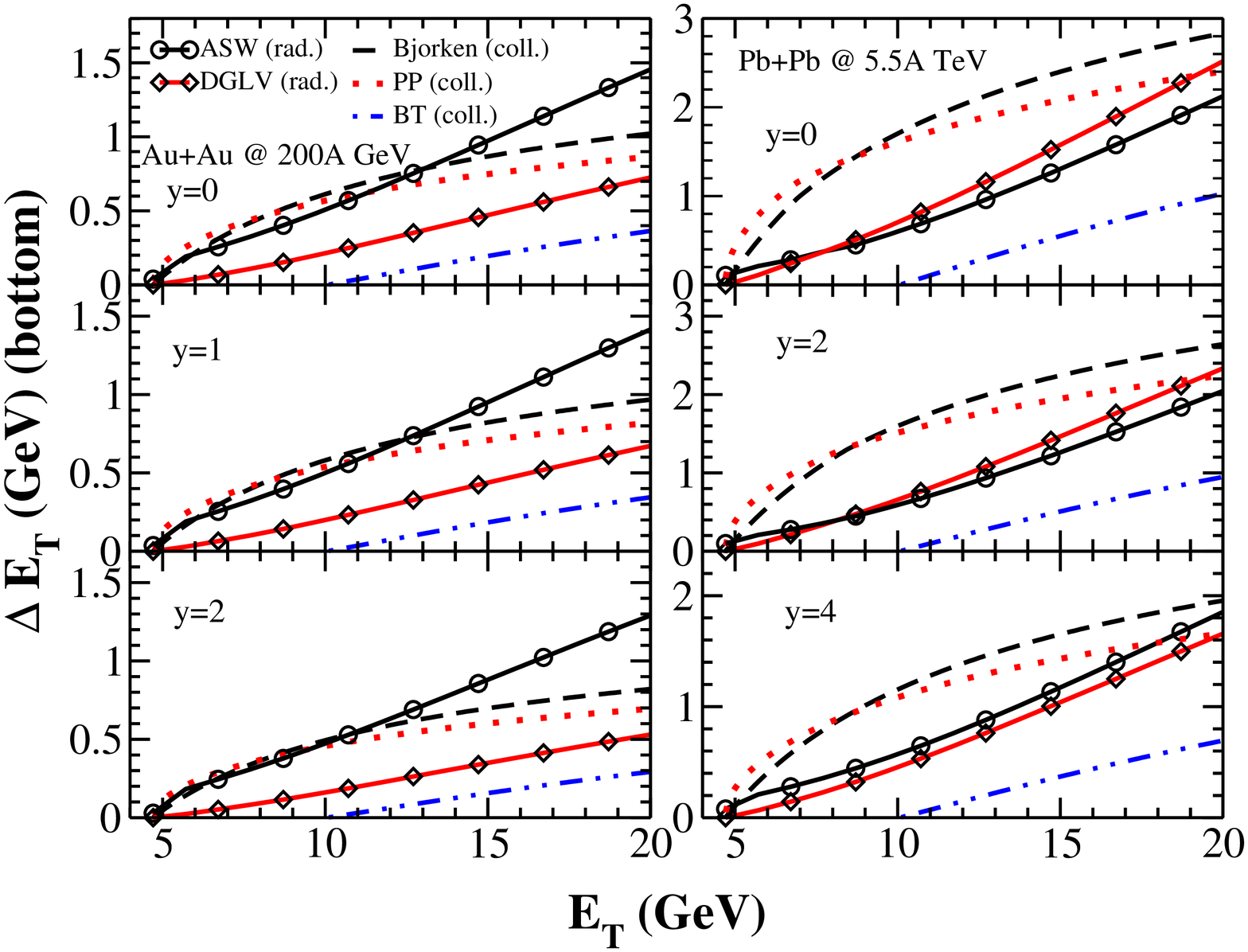}
\caption{\label{2} Same as Fig.~\ref{eloss-c} for a bottom 
quark}
\label{eloss-b}
\end{center}
\end{figure}

For 
the calculation of radiative energy 
loss, we consider the treatment of Djordjevic, Gyulassy, Levai, and Vitev 
(DGLV)~\cite{dglv1, dglv2} using opacity expansion, 
the treatment of Armesto, Salgado, and Wiedemann (ASW)~\cite{asw}
 using path 
integral formalism for medium-induced gluon radiations off massive quarks, 
and the treatment of Xiang, 
Ding, Zhou, and 
Rohrich (XDZR)~\cite{xdzr} using light cone path integral approach. 
Detailed expressions for these formalisms are given in the Appendix B.

\section{Results for Energy Loss}

We compare the results for transverse 
energy loss for a heavy quark 
using these different energy loss treatments for several rapidities. 
We plot the transverse energy loss of charm and bottom quarks, 
$\Delta E_T$ as a function of transverse energy $E_T$ 
$(\sqrt{p_T^2+M^2})$ in Figs.~\ref{eloss-c} 
and ~\ref{eloss-b}, at RHIC and LHC energies. 

Several interesting features emerge. We see that the collisional energy loss 
for charm quarks at RHIC and LHC energies is only marginally dependent 
on the rapidity and the BT formalism gives largest energy loss, as expected.
 In our treatment, change of rapidity implies a change in the 
temperature of the plasma. Thus these results suggest a weaker dependence on 
the temperature and the average path length for the energy loss suffered by 
charm quarks due to collisions. 

The radiative energy loss, on the other hand, shows a much more complex 
behaviour and is quite different for the different formalisms under 
consideration. We note that the ASW formalism for radiative energy loss gives 
largest degradation in the energy at all rapidities (except for $E_T\,<\,5\,
{\rm{GeV}}$ at LHC, where it is comparable to the collisional energy loss). We 
also see that the DGLV and the XDZR formalisms give similar results at RHIC 
energies, at the three rapidities under consideration. On the other hand, at 
LHC energy, the ASW and DGLV formalisms provide nearly identical results for 
energy loss for charm quarks at y=0 and y=2, and the corresponding results at 
y=4 differ by about 10$\%$. This, we feel, is due to a more complex dependence 
on the average path length in the ASW formulation.

The collisional energy loss for bottom quarks using the PP and the Bjorken's 
formulation are seen to be quite similar at RHIC and LHC energies for all 
rapidities under consideration. The BT formulation due to the neglect of the 
u-channel, gives a much smaller energy loss, both for RHIC and LHC energies 
and at all rapidities.

We have already mentioned that due to the 
numerical approximations used, the XDZR 
formulation is not valid for evaluation of the radiative energy loss for bottom 
quarks. The ASW and DGLV radiative energy loss formalisms
 show a more complex dependence on the mass and 
the average path length. 
The ASW formulation gives a larger energy loss at RHIC energy, though the
 results are again comparable at LHC energy at all rapidities. 
We note that while the collisional and radiative energy losses
 for bottom quarks 
at RHIC energy are comparable, the 
collisional energy loss dominates over the radiative energy loss in the 
$E_T$ range under consideration at LHC energy. 
We have confirmed that at higher $E_T$, the radiative energy loss again starts 
dominating.

This rich structure suggests that description of energy loss for one (quark) 
mass at one rapidity, and one energy
 may not be enough to identify the most reliable 
treatments, for this.
\section{$R_{AA}$ for heavy quarks}

The nuclear modification factor $R_{AA}$ for heavy quarks can be 
expressed as:
\begin{eqnarray}
R_{AA}(b)=\frac {dN^{AA}/d^2p_T\,dy}{T_{AA}(b)\,d \,
 \sigma^{NN}/d^2p_T\,dy},
\end{eqnarray}
where, as mentioned earlier, 
 $T_{AA}(b)$ is the nuclear overlap function for impact parameter b, 
calculated using Glauber model. We get $T_{AA}$$\approx$ \,280 fm$^{-2}$
 for Au+Au collisions at RHIC and $\approx$ 290 fm$^{-2}$ for Pb+Pb 
collisions at LHC, for b\,=\,0 fm.

\begin{figure}[ht]
\begin{center}
\includegraphics[width=14 pc,angle=-90]{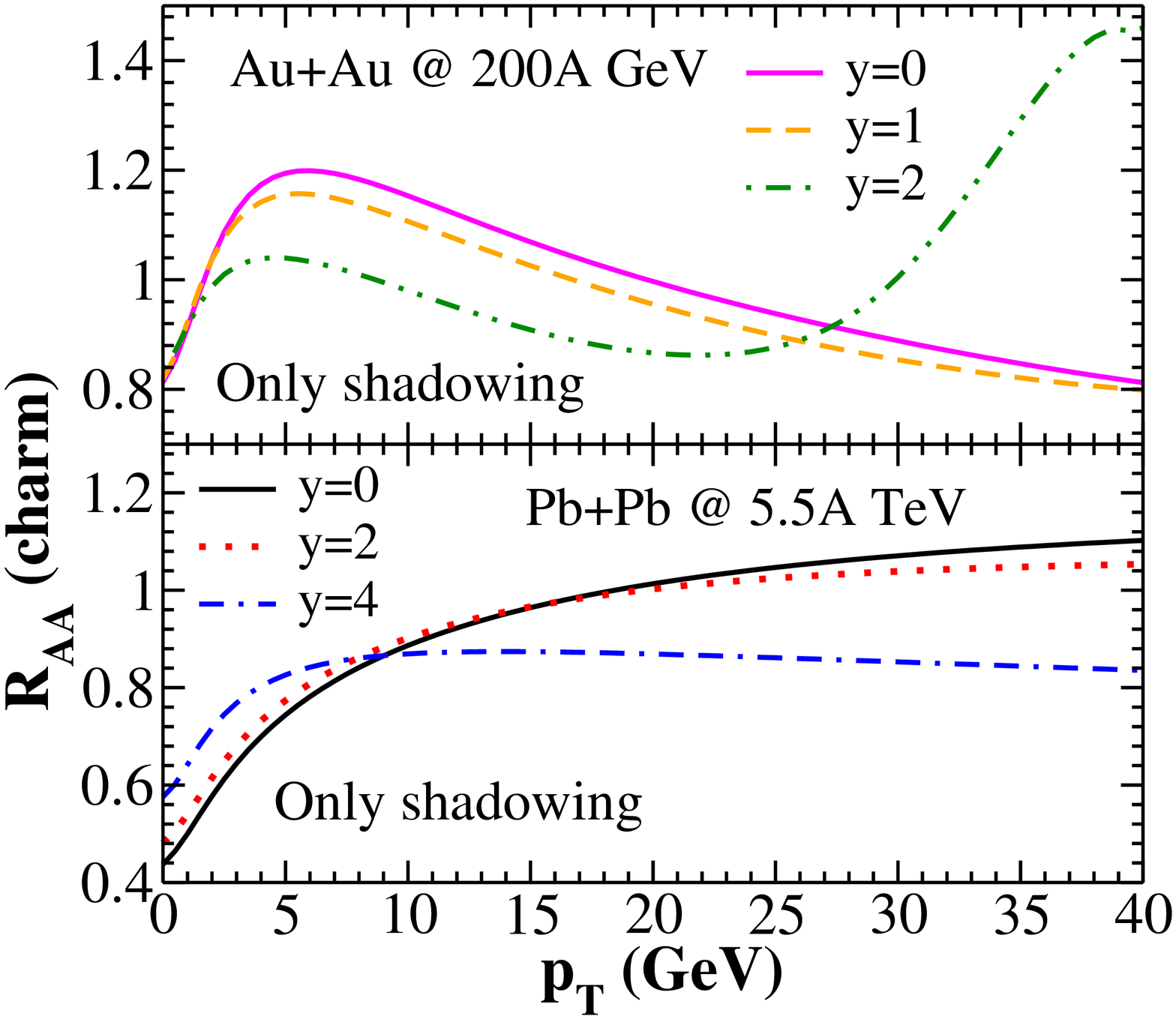}
\includegraphics[width=14 pc,angle=-90]{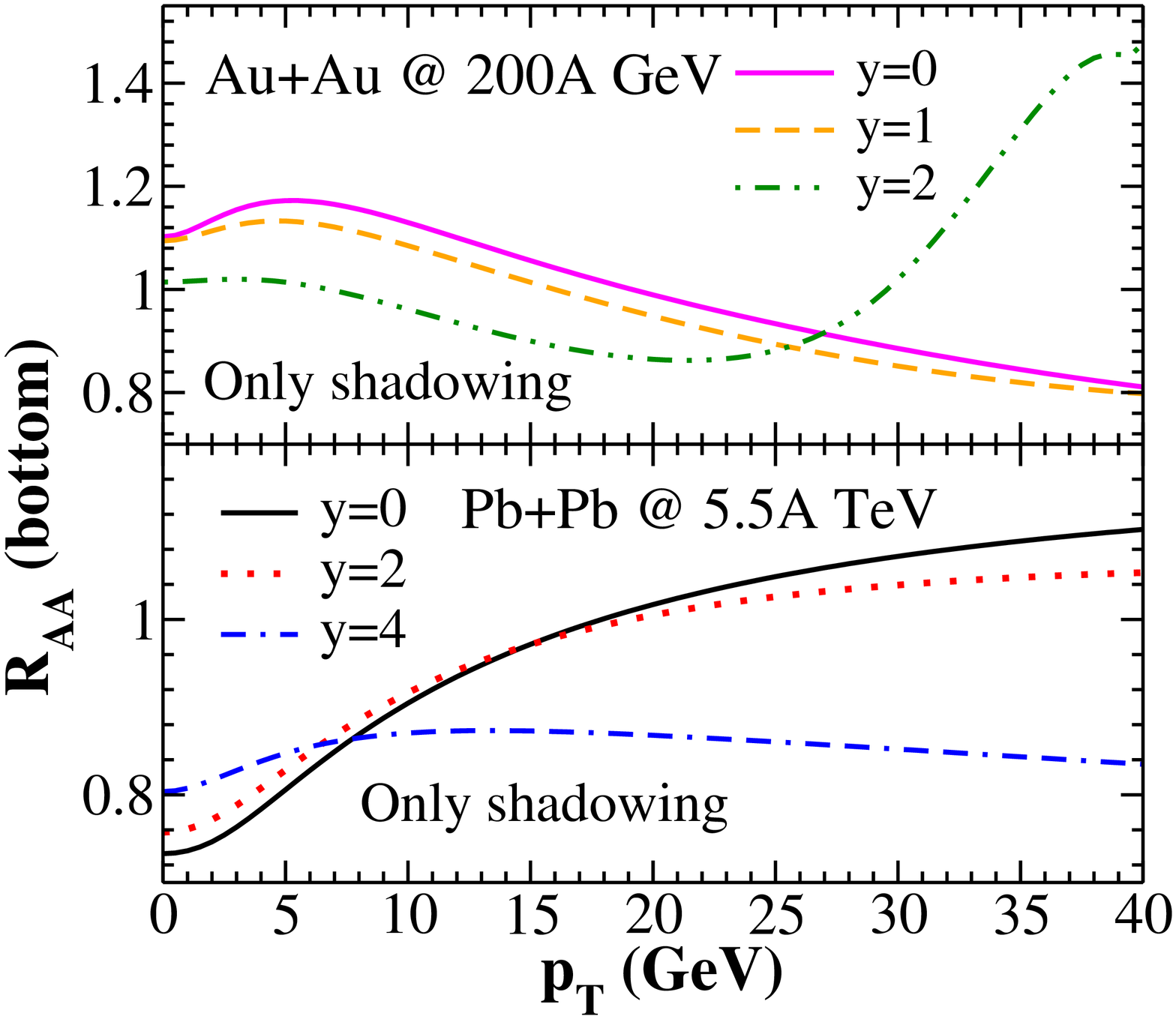}
\caption{\label{2}[Left panel] $R_{AA}$ of charm quarks with only 
nuclear shadowing effect at more forward rapidities. 
[Right panel] Same as left panel for bottom quark.}
\label{raaws}
\end{center}
\end{figure}

As a first step, we give the results for $R_{AA}$ with only nuclear 
shadowing effect for production of charm and 
bottom quarks at the rapidities considered earlier for RHIC and LHC energies 
(see Fig.~\ref{raaws}). We see that the 
fairly large masses of the charm and bottom 
quarks, the kinematics, and the rich behaviour of the structure function 
with x and $Q^2$, lends interesting features to $R_{AA}$. 

We see that $R_{AA}$ 
for charm and bottom quarks for y\,=\,0 and y\,=\,1 are quite similar at RHIC 
energy. Similarly, the results for y\,=\,0 and y\,=\,2 are only marginally 
different at LHC energy. The results at larger y are more strongly affected 
due to increased variation in the 'x' values (see Eq.~\ref{x1}) which 
contribute. In order to do a full justice to these interesting results, we 
now discuss them individually.

For charm quarks at RHIC energy, we see a suppression at lower $p_T$, an 
enhancement at intermediate $p_T$, and again a suppression at larger $p_T$, 
for y\,=\,0 and y\,=\,1, while for y\,=\,2, $R_{AA}$ starts at about 0.8, 
goes up to a value slightly more than 1, then drops again to about 0.8 at 
$p_T$\,$\approx$\,20 GeV, and rise again to beyond 1 at 
$p_T$\,$\approx$\,40 GeV. Since the energy loss of the charm quarks always 
rises with increasing $p_T$, this would introduce interesting features in 
$R_{AA}$ after this is accounted for, unless of course the $p_T$ spectrum 
for the quarks drops too rapidly. We shall come back to this point again.
 The increased energy at LHC then provides a larger suppression at low $p_T$ 
for all the rapidities. In an interesting development, $R_{AA}$ for 
y\,=\,0 and y\,=\,2 rises beyond 1 at $p_T$\,$\approx$\,20 GeV, while it stays 
below 1 up to $p_T$\,$\approx$\,40 GeV, for y\,=\,4.

The results for bottom quarks are even more interesting. Due to the large mass 
of the bottom quarks, at RHIC energy, the $R_{AA}$ for lower $p_T$ for 
y\,=\,0 and y\,=\,1 is already starts getting contributions from the region 
of x where anti-shadowing appears. Thus $R_{AA}$ starts at a value which is
 more than 1 at lower $p_T$, goes up, up to $p_T$\,$\approx$\,5 GeV and then 
drops again. For y\,=\,2 on the other hand, it starts at a value close to 1, 
drops by about 10\,$\%$ at $p_T$\,$\approx$\,20 GeV and rises again. At LHC 
energy, $R_{AA}$ for bottom quarks for y\,=\,0 and y\,=\,2 starts at 
$\approx$\,30\,$\%$ below 1 and then rises steadily to about 1.1 at 
$p_T$\,$\approx$\,40 GeV. The results at y\,=\,4 remain close to 0.8, rising 
slightly at intermediate $p_T$.
\begin{figure}[ht]
\begin{center}
\includegraphics[width=14 pc,angle=-90]{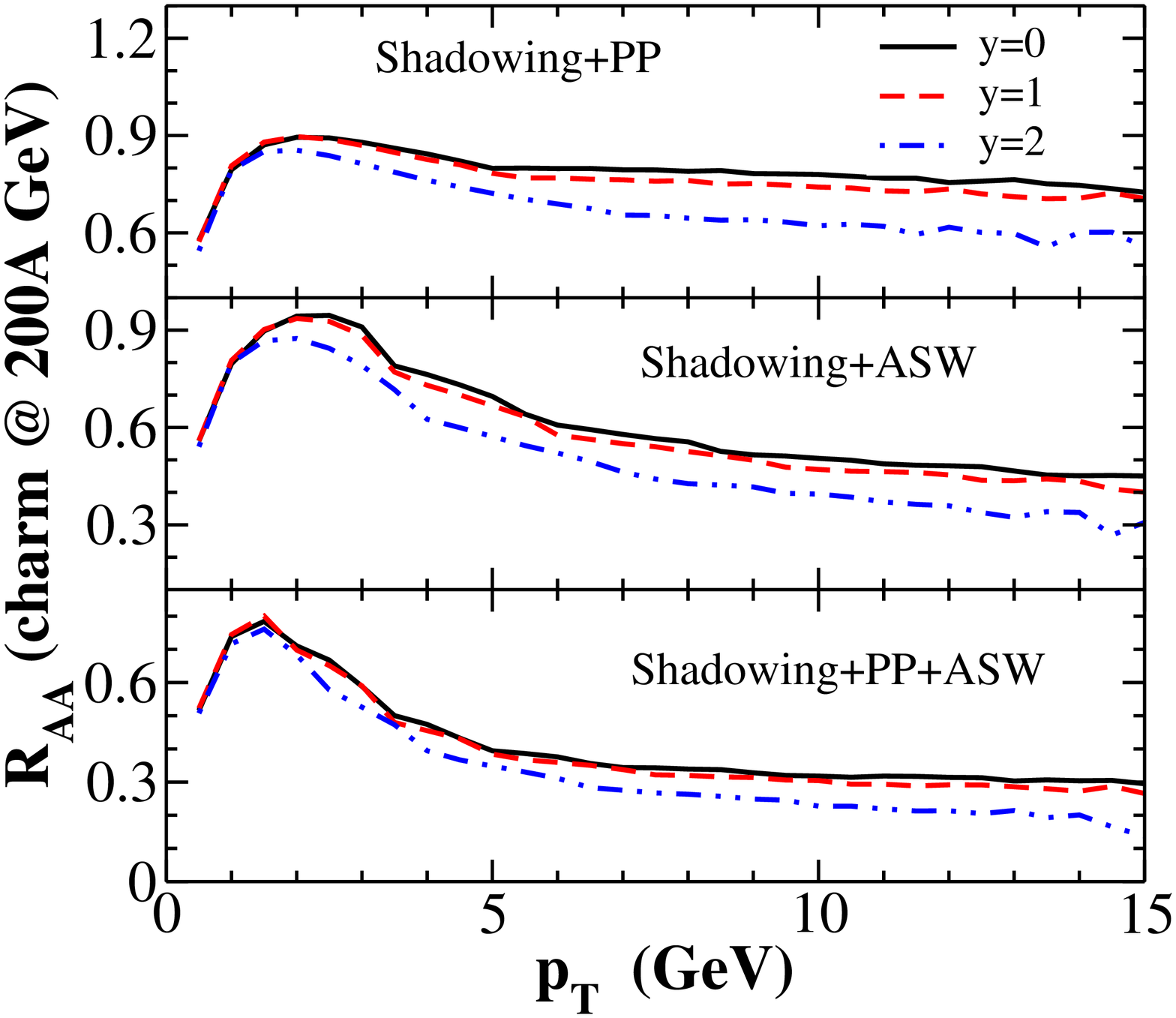}
\includegraphics[width=14 pc,angle=-90]{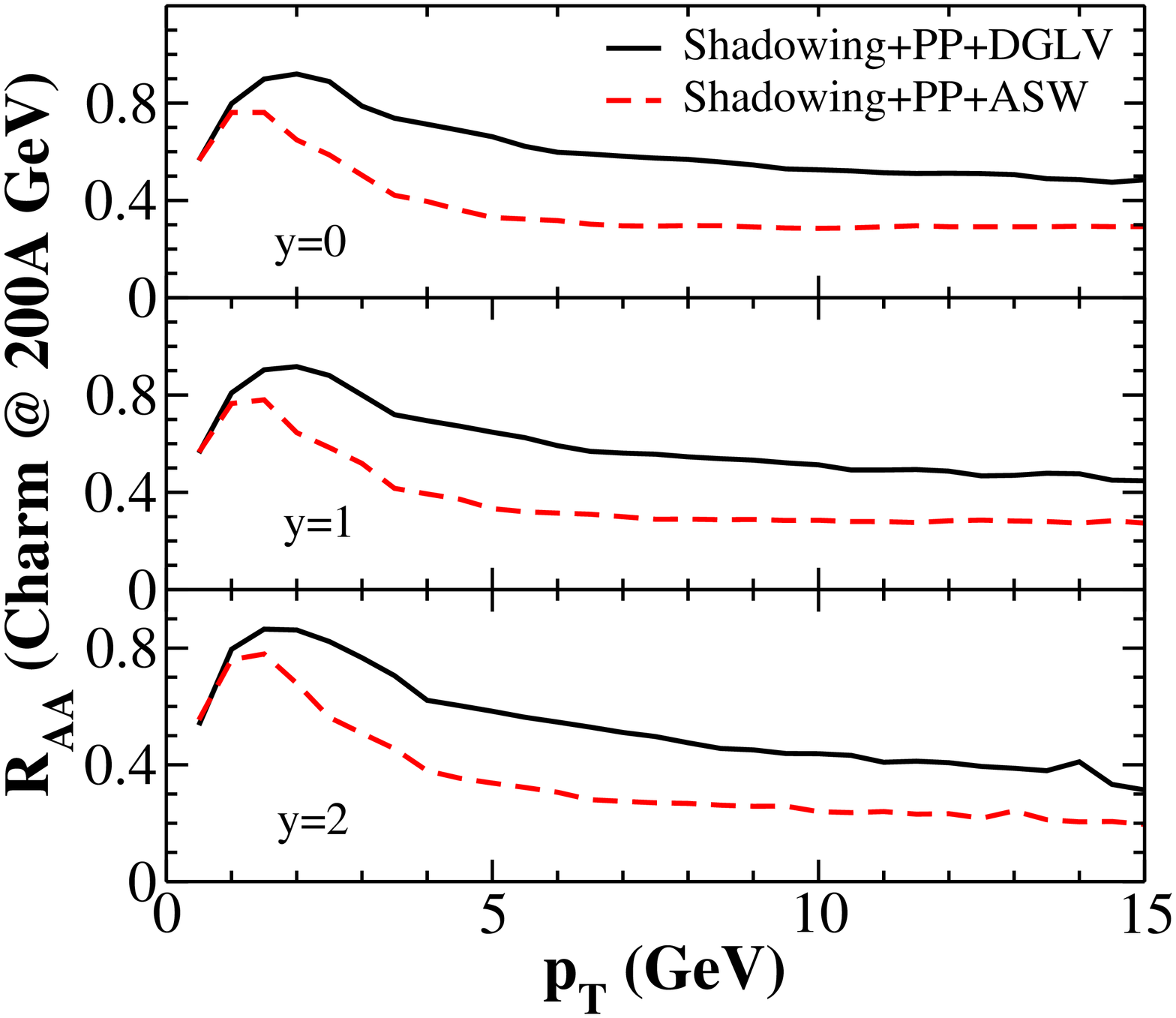}
\caption{\label{2} [Left panel] $R_{AA}$ of charm quarks with the 
nuclear shadowing effect as well as the energy loss 
at more forward rapidities at RHIC energy.
[Right panel] Comparison of the relative nuclear suppression using ASW 
and DGLV formalisms for charm quarks at different rapidities at 
RHIC energy.}
\label{raac-rhic}
\end{center}
\end{figure}

Now let us discuss our results for $R_{AA}$ with the additional inclusion 
of collisional and radiative energy losses. We shall restrict our 
consideration to inclusion of collisional energy loss using the PP formulation 
and the radiative energy loss using ASW or DGLV formulation.

Fig.~\ref{raac-rhic} gives our findings for charm quarks at RHIC energy. We 
see that the outcome of shadowing and energy loss gives an interesting 
structure to $R_{AA}$, as expected. We see that the final $R_{AA}$ starts at 
about 40\,$\%$ below 1, goes up to about 0.8 at $p_T$\,$\approx$\,2 GeV, and 
then drops to a value of $\approx$\,0.3 at $p_T$\,$\approx$\,15 GeV. In an 
interesting development, we see that the combination of the shadowing and 
energy loss gives a marginally larger suppression at y\,=\,2 compared 
to y\,=\,0, even though the fractional energy loss is higher at smaller y (see 
Fig.~\ref{eloss-c}). We have also given a comparison of $R_{AA}$ by replacing 
the ASW formulation for the radiative energy loss with the DGLV treatment, and 
see that the former gives a larger suppression at all y (see 
also Fig.~\ref{eloss-c}). We recall that the single electrons produced from 
the semi-leptonic decay of charm mesons~\cite{info3} show $R_{AA}$ of about 
0.2\,$\sim$\,0.3 for $p_T$\,$>$\,2 GeV, in a very encouraging agreement with 
these results. This will be pulbished shortly.
\begin{figure}[ht]
\begin{center}
\includegraphics[width=17 pc,angle=-90]{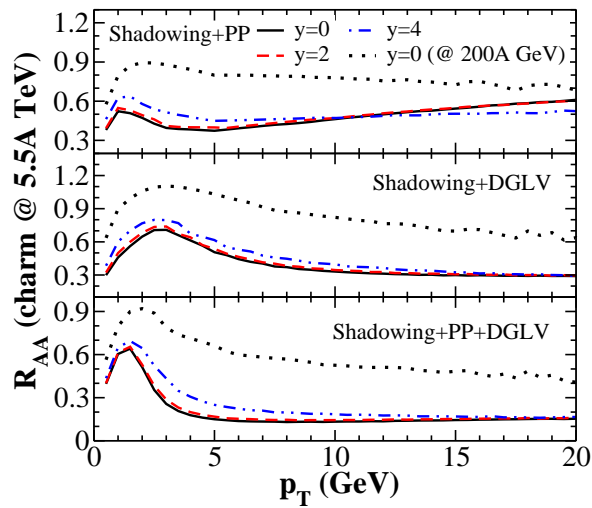}
\caption{\label{2} $R_{AA}$ of charm quarks with the 
nuclear shadowing effect as well as the energy loss 
at more forward rapidities at LHC energy.}
\label{raac-lhc}
\end{center}
\end{figure}

The results for the $R_{AA}$ for charm quarks at LHC energy are shown in 
Fig.~\ref{raac-lhc}. We show the results only for the DGLV formulation for 
the radiative 
energy loss, as we have seen that it is quite similar to that for the 
ASW formalism for charm quarks at LHC energy. We see roughly similar behaviour, 
in that the $R_{AA}$ starts from a lower value at $p_T$\,$\approx$\,0 GeV, 
rises up to $p_T$\,$\approx$\,2 GeV, and then drops to a level of about 0.2 
at larger $p_T$. We also find a marginally larger suppression for y\,=\,0 
compared to that for y\,=\,4. The results for y\,=\,0 for RHIC energy are also 
given for a comparison which suggests a much larger suppression at LHC, as 
expected.
\begin{figure}[ht]
\begin{center}
\includegraphics[width=14 pc,angle=-90]{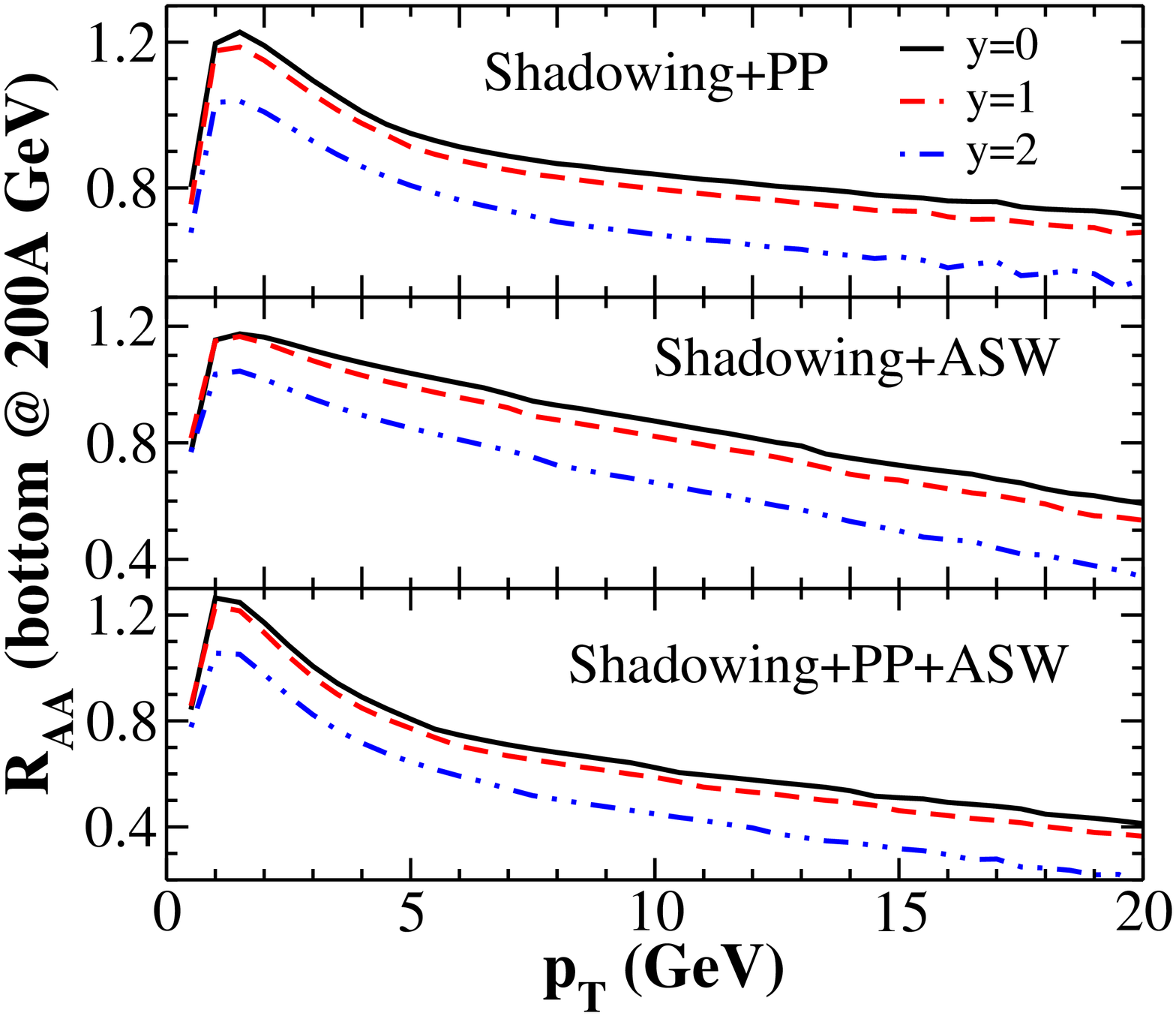}
\includegraphics[width=14 pc,angle=-90]{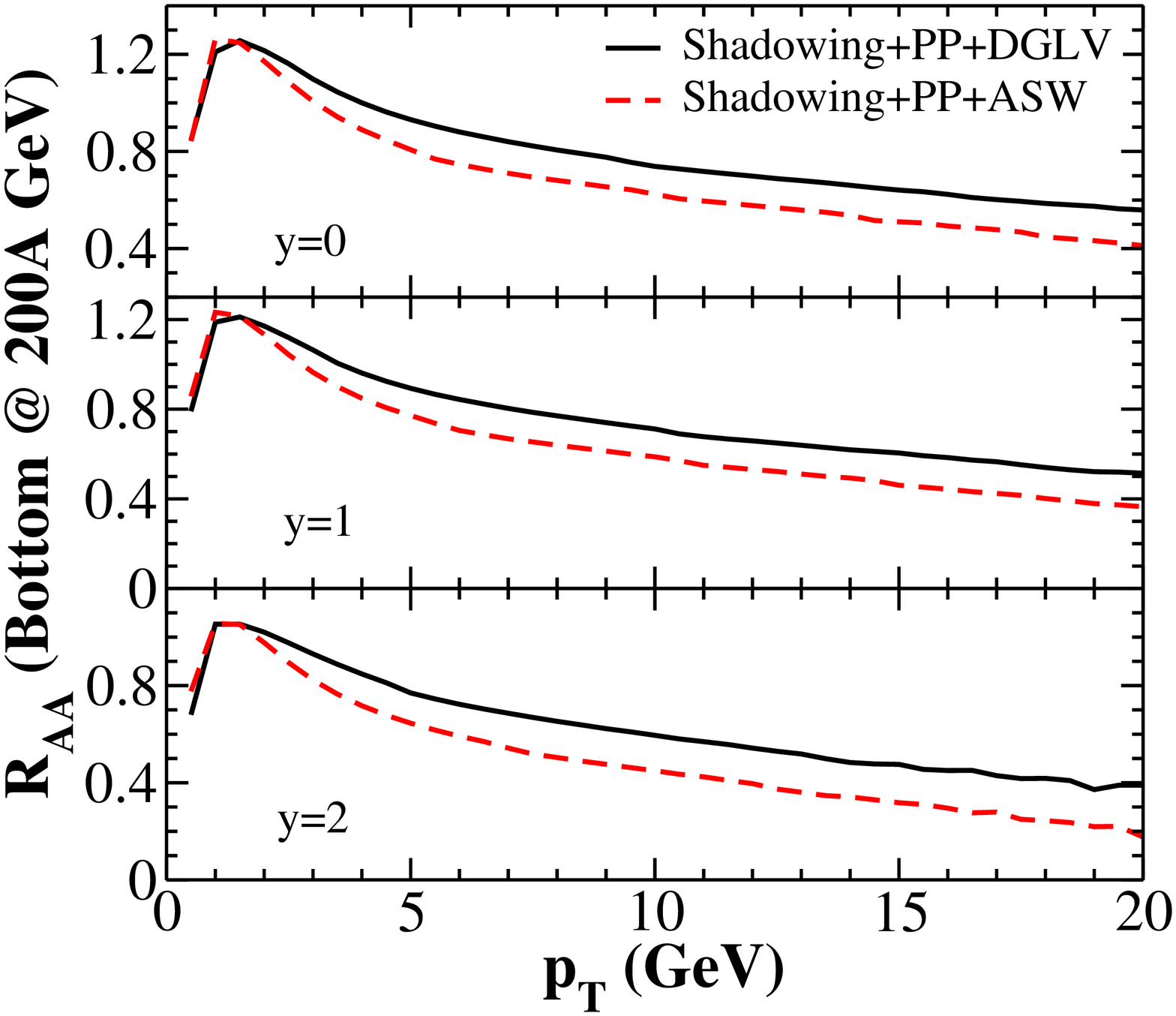}
\caption{\label{2} Same as Fig.~\ref{raac-rhic} for bottom quarks.}
\label{raab-rhic}
\end{center}
\end{figure}

Next we discuss our findings for nuclear suppression for bottom quarks at RHIC 
energy (see Fig.~\ref{raab-rhic}). The shadowing and the large mass of the 
bottom quarks, with its consequences, gives an $R_{AA}$\,$\approx$\,0.8 at 
$p_T$\,$\approx$\,0 GeV, which goes up to about 1.2 at $p_T$\,$\approx$\,2 GeV,
 and then drops to about 0.3\,$\sim$\,0.4 at larger $p_T$. The shadowing 
results in a larger suppression for y\,=\,2 compared to y\,=\,0 (see Fig.~
\ref{raaws}), even though the energy loss is slightly lower for larger y 
(see Fig.~\ref{eloss-b}). Results obtained by replacing the ASW formulation 
with the DGLV treatment show a smaller suppression, as for charm 
quarks (see Fig.~\ref{raac-rhic}).
\begin{figure}[ht]
\begin{center}
\includegraphics[width=17 pc,angle=-90]{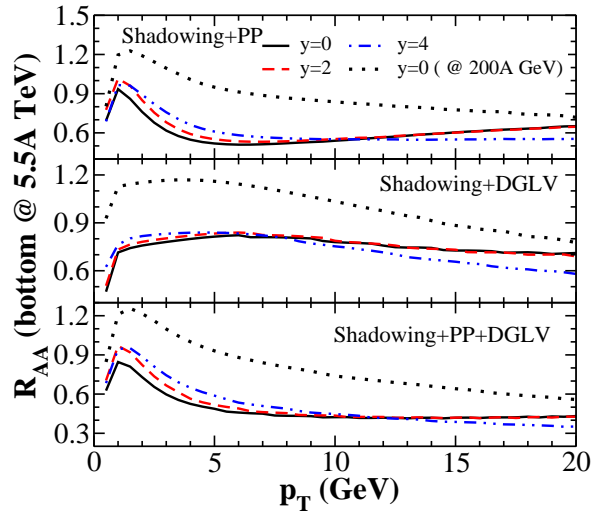}
\caption{\label{2} Same as Fig.~\ref{raac-lhc} for bottom quarks.}
\label{raab-lhc}
\end{center}
\end{figure}

Finally in Fig.~\ref{raab-lhc} we have given our results for $R_{AA}$ 
for bottom quarks at LHC energy for y\,=\,0, 2 and 4 using shadowing, 
collisional energy loss and radiative energy loss using the DGLV treatment. 
The results using ASW treatment are expected to be quite similar as seen from 
Fig.~\ref{eloss-b}. The results for y\,=\,0 at RHIC energy are also given for 
a ready reference. We again see a trend which is common to our results that 
$R_{AA}$ starts at $\approx$\,0.6 at $p_T$\,$\approx$\,0 GeV, goes up to 
 $\approx$\,1.1 at $p_T$\,$\approx$\,2 GeV, and then slowly drops to about 
0.4 at $p_T$\,$\approx$\,20 GeV. The overall effect of shadowing and energy 
loss is seen to lead to very similar values for $R_{AA}$ from y\,=\,0 to
 y\,=\,4. Of course the $R_{AA}$ for RHIC energy is about twice as large,
 showing
 a much reduced suppression.

\section{Summary and Discussion}

We have made a detailed study of charm and bottom production from initial 
fusion of partons in relativistic collision of heavy nuclei. As a first step 
we have checked the usefulness of lowest order pQCD to reproduce the NLO 
results for the $p_T$ distribution of charm and bottom quarks in pp collisions.
 Next we have checked our predictions against experimental results of the charm 
and bottom quarks production from such collisions.

We have further obtained the average energy loss suffered by heavy quarks 
at several rapidities for RHIC and LHC energies due to collisions and gluon 
radiations. We have obtained the nuclear suppression factor $R_{AA}$ by 
additionally incorporating nuclear shadowing for the cases under study. A rich 
picture of dependence of $R_{AA}$ on y, $p_T$, incident energy, and the mass 
of the heavy quarks emerges. We have noted that our findings would support 
the suppression of single electrons seen at RHIC.

Before concluding, we discuss some of the short-comings of the present work. 
These initial calculations can be improved in several ways. We have used 
(1\,+\,1) dimensional Bjorken hydrodynamics and assumed it to apply at all y. 
Since the heavy quarks will loose most of their energy at very early times, 
this may not be a serious short-coming. Still we are looking at the possibility 
of using a full fledged (3\,+\,1) dimensional hydrodynamics calculations, also 
at b\,$\ne$\,0. We are incorporating the single electron decay of the
 resulting D and B mesons, along with the results for their back to back 
correlation. We expect this to be rewarding, especially in conjunction with 
NLO results for pp collisions, as it may throw up an interesting detail about 
differences of NLO results and the results with energy loss. These will be 
published shortly.

Finally, we conclude that the description for energy loss for one quark mass 
at one rapidity for a particular incident energy may not be 
sufficient to identify the most reliable energy loss treatment for either 
collisional or radiative energy loss valid for all cases. 

\section*{Acknowledgments} 
We gratefully acknowledge the use of MNR-NLO code~\cite{mnr} made available 
to us by the authors. We also thank M. G. Mustafa for useful discussions.

\section*{Appendix}

\appendix

\section{Collisional energy loss}

\subsection*{A.1. Bjorken}

Bjorken argued that the elastic energy loss by partons in the QGP is very 
similar to the energy loss due to ionization due to passage of 
charged particles in ordinary matter~\cite{bjorken}. This treatment was
 adapted by Braaten and Thoma for heavy quarks~\cite{bt1}. 
The fractional energy loss suffered by the heavy quark 
due to collisions with the quarks and gluons given as: 

\begin{eqnarray}
\frac{dE}{dx}=
\frac{8\,\pi\, \alpha_s^2\,T^2}{3}\left(1\,+\,\frac{N_f}{6}\right)
\left[\frac{1}{v}\,-\,\frac{1\,-\,v^2}{2\,v^2}\log\frac{1\,+\,v}{1\,-\,v}\right]
\log\frac
{q_{\rm {max}}}{q_{\rm{min}}}, 
\label{bjorken}
\end{eqnarray}
where $v$ is the velocity of the heavy quark. As suggested by Braaten and 
Thoma we use the upper 
limit of the momentum transfer $q_{\rm {max}}$ as $\sqrt{4\,T\,E}$ 
and the lower limit
 of the momentum transfer $q_{\rm {min}}$ as $\sqrt{3\,m_g}$.

 The thermal gluon mass $m_g$ can be expressed as $m_g$=$\mu/\sqrt{2}$ where 
$\mu$=$\sqrt{4\,\pi\,\alpha_s\,T^2\,\left(1\,+\,\frac{N_f}{6}\right)}$ 
is the Debye screening mass.

\subsection*{A.2. Braaten and Thoma}

Braaten and Thoma first developed a theoretical formalism 
to find the collisional energy loss of a 
muon propagating 
through a plasma of electrons, positrons and photons to leading order in QED
~\cite{bt3}. This work was further extended by them to
 calculate the collisional energy loss of heavy quarks propagating through QGP
~\cite{bt1}. The energy loss formulation is given in two energy 
regimes:
The fractional collisional energy loss of heavy quarks with energy $E<<M^2/T$ is

\begin{eqnarray}
\frac{dE}{dx}=
\frac{8\,\pi \,\alpha_s^2\,T^2}{3}\left(1\,+\,\frac{N_f}{6}\right)
& &\left[\frac{1}{v}-\frac{1\,-\,v^2}{2\,v^2}\log\frac{1\,+\,v}{1\,-\,v}
\right]\times 
\nonumber \\& &
\log
\left(2^{\frac{N_f}{6\,+\,N_f}} \, B(v) \, \frac{E\,T}{m_g\,M}\right) 
\label{bt1}
\end{eqnarray}

The fractional 
collisional energy loss of heavy quarks with energy $E>>M^2/T$ is

\begin{eqnarray}
\frac{dE}{dx}=
\frac{8\,\pi \,\alpha_s^2\,T^2}{3}\left(1\,+\,\frac{N_f}{6}\right)
\log\left(2^{\frac{N_f}{2\,\left(6\,+\,N_f\right)}} \, 0.92 \,
 \frac{\sqrt{E\,T}}{m_g}\right)
, 
\label{bt2}
\end{eqnarray} 
where B($v$) is a smooth function of velocity having value in the range 
0.6\,-\,0.7.  
Braaten and Thoma have shown the crossover energy between these
 energy regimes as $E_{\rm{cross}}\, =\,1.8 \,\times \, M^2/T$ for $N_f$\,=\,2.

\subsection*{A.3. Peigne and Peshier}

In BT formalism, it was 
assumed that the momentum exchange in the elastic 
scattering process is much less than the energy carried by the heavy quarks.
Peigne and Peshier pointed out that this assumption is not reliable in the 
energy regime $E\,>>\,M^2/T$, and corrected it in the QED case while calculating 
 the collisional energy loss of a muon in QED plasma~\cite{pp1}. This work in 
QED is then used by them to derive the collisional energy loss 
suffered by heavy quarks while passing through QGP~\cite{pp}.
 
The fractional collisional energy loss suffered by heavy quarks as proposed by 
Peigne and Peshier is

\begin{eqnarray}
\frac{dE}{dx}=
\frac{4\,\pi \,\alpha_s^2\,T^2}{3}\left[\left(1\,+\,\frac{N_f}
{6}\right)
\log\frac{E\,T}{\mu^2}+\frac{2}{9}\log \frac{E\,T}{M^2}\,+\,
c\left(N_f\right)\right] 
\label{pp}
\end{eqnarray}
and $c\left(N_f\right)\,\approx \,0.146 \,N_f\,+\,0.05$.

\section{Radiative energy loss}
\subsection*{B.1. Djordjevic, Gyulassy, Levai, and Vitev}

For massless quarks, Gyulassy, Levai and Vitev (GLV) calculated the induced 
radiation to arbitrary order in opacity $\chi^n$ ($\chi\,=\, L/\lambda$) 
of the 
plasma~\cite{glv} where $\lambda$ is the mean free path of the quark. 
In Djordjevic, Gyulassy, Levai, 
and Vitev (DGLV) formulation~\cite{dglv1}, the 
GLV method is generalized to estimate the first 
order induced radiative energy loss including the kinematic effect for 
heavy quarks. Wicks {\it et al.}~\cite{dglv2} present a simplified form of 
the DGLV formalism for the average radiative energy loss of heavy quarks:

\begin{eqnarray}
\Delta E=\frac{c_F\,\alpha_s}{\pi}\frac{E\,L}{\lambda_g}
 \int\limits_{\frac{m_g}{E\,+\,p}}^{1\,-\,\frac{M}{E\,+\,p}}dx\int\limits_{0}^{\infty}
\frac{4\,\mu^2\,q^3\,dq}{\left(\frac{4\,E\,x}{L}\right)^2\,+\,
\left(q^2\,+\,\beta^2\right)^2}
\left(A\log B+C\right),
\label{dglv}
\end{eqnarray}
where
\begin{eqnarray}
\beta^2 = m_g^2\,(1\,-\,x)\,+\,M^2\,x^2,\nonumber\\
\frac{1}{\lambda_g}\,=\,\rho_g\,\sigma_{gg}\,+\,\rho_q\,\sigma_{qg},\nonumber\\
\sigma_{gg}=\frac{9\,\pi\,\alpha_s^2}{2\mu^2},\nonumber\\
\sigma_{qg}=\frac{4}{9}\,\sigma_{gg},\nonumber\\
\rho_g=16 \, T^3 \,\frac{1.202}{\pi^2},\nonumber\\
\rho_q=9 \, N_f \, T^3\, \frac{1.202}{\pi^2},\nonumber\\
A=\frac{2\,\beta^2}{f_{\beta}^3}\left(\beta^2\,+\,q^2\right),\nonumber\\
B=\frac{\left(\beta^2\,+\,K\right)\left(\beta^2\,Q_\mu^-\,+\,Q_\mu^+\,
Q_\mu^+\,+\,Q_\mu^+\,f_\beta
\right)}
{\beta^2\left(\beta^2\left(Q_\mu^-\,-\,K\right)\,-\,Q_\mu^-K\,+\,Q_\mu^+\,
Q_\mu^+\,+\,f_\beta\, f_\mu\right)}, \nonumber \\
C=\frac{1}{2\,q^2\,f_{\beta}^2\,f_{\mu}}\left[\right.
\beta^2\,\mu^2\left(2\,q^2\,-\,\mu^2\right)
\,+\,\beta^2\left(\beta^2\,-\,\mu^2\right)K\,+\,Q_\mu^+\left(\beta^4\,-\,
2\,q^2\,Q_\mu^+\right)
 \nonumber \\
\ \ \ \ \ \ \ \ \ \ \ \ \ \ \ \ \ 
\,+\,f_\mu\left(\beta^2\left(\,-\,\beta^2\,-\,3q^2\,+\,\mu^2\right)\,+\,2\,
q^2\,Q_\mu^+\right)
\,+\,3\beta^2\,q^2\,Q_k^-\left.\right], \nonumber \\
K=\left(2\,p\,x(1\,-\,x)\right)^2,\nonumber \\
Q_\mu^{\pm}=q^2\,\pm\,\mu^2,\nonumber \\
Q_k^{\pm}=q^2\,\pm \,K,\nonumber \\
f_\beta=f\left(\beta,\, Q_\mu^-,\,Q_\mu^+\right), \nonumber \\
f_\mu=f\left(\mu, \,Q_k^+,\,Q_k^-\right) \nonumber \\
{\rm {and}} \nonumber\\
f(x,\, y ,\,z)=\sqrt{x^4\,+\,2\,x^2\,y\,+\,z^2}.
\end{eqnarray}

\subsection*{B.2. Armesto, Salgado, and Wiedemann}

In Armesto, Salgado, and Wiedemann (ASW), formulation~\cite{asw} 
path integral method for medium-induced 
gluon radiation is employed to calculate the radiative energy loss of 
heavy quarks. This formalism provides the analysis of the double 
differential medium-induced gluon distribution by the heavy quarks as a 
function of transverse momentum. 
The average radiative energy loss is

\begin{eqnarray}
\Delta\, E=\frac{\alpha_s\,c_F}{\pi}\left(2\,n_0\,L\right)&&
\int\limits_0^Ed\omega\int\limits_0^\frac{R}{2\gamma^2}dk^2 \int\limits_0^\infty dq^2 \times \nonumber \\
&&\frac{\left(q^2\,+\,\bar{M}^2\right)\,-\,\frac{1}{\gamma}\,\sin\left[\gamma
\left(q^2\,+\,\bar{M}^2\right)\right]}{\left(q^2\,+\,\bar{M}^2\right)^2}
\nonumber \\
& &\times \frac{q^2}{q^2\,+\,\bar{M}^2}\nonumber\\ & & \, \times
\frac{\left(k^2\,+\,\bar{M}^2\right)\,+\,\left(k^2\,-\,\bar{M}^2\right)
\left(k^2\,-\,q^2\right)}
{\left(k^2\,+\,\bar{M}^2\right)\left[\left(1\,+\,k^2\,+\,q^2\right)^2\,-\,4\,
k^2\,q^2\right]^\frac{3}
{2}}.
\label{asw}
\end{eqnarray}
In the above equation the gluon energy, transverse momentum and heavy 
quark mass are expressed as 
dimensionless parameters. The rescaled dimensionless parameters:
\begin{eqnarray}
\bar{M}^2 \,
\equiv\,\frac{1}{2}\,\left(\frac{M}{E}\right)^2\,\frac{R}{\gamma^2}, 
\nonumber \\
R=\omega_c\,L, \, \omega_c\,\equiv \,\frac{1}{2}\,\mu^2\,L, \,
  {\rm {and}}  \,  \gamma \,\equiv\,\frac{\omega_c}{\omega}.
\end{eqnarray}
We use the parameter $\rm{n_0\,L}$\,=\,4.
\subsection*{B.3. Xiang, Ding, Zhou, and Rohrich}
 In Xiang, 
Ding, Zhou, and 
Rohrich (XDZR) formulation~\cite{xdzr}, light 
cone-path integral method is used to calculate the gluon radiation from 
heavy quarks where an analytical expression is obtained for the heavy 
quark radiative energy loss. 

The average radiative energy loss is
\begin{eqnarray}
\Delta E=\frac{\alpha_s\,c_F}{4}\,\frac{L^2\,\mu^2}{\lambda_g}& &\left[
\log\frac{E}{\omega_{cr}}\,+\,\frac{m_g^2\,L}{3\,\pi\,\omega_{cr}}\left(1\,-\,
\frac{\omega_{cr}}{E}\log\frac{E^2}{2\mu^2\,L\,\omega_{cr}}\,+\,
\log\frac{\omega_{cr}}{2\,\mu^2\,L}\right)\right. \nonumber \\
& & \left.\,+\,\frac{M^2L}{3\,\pi \,E}\left(\frac{\pi^2}{6}-\frac
{\omega_{cr}}{E}\log\frac{\omega_{cr}}{2\,\mu^2\,L}\,+\,
\log\frac{E}{2\,\mu^2\,L}\right) \right],
\label{xdzr}
\end{eqnarray}
where $\omega_{cr}$\,=\,2.5 GeV. This 
analytical expression is derived by using an expansion for Bessel function
~\cite{xdzr}, 
which is valid only for not too large mass of the quarks. 
We use it only for charm quark.

\end{document}